\documentclass[12pt,letter]{article}

\usepackage{graphicx, epsfig, color, cite}
\def\to{\rightarrow}

\def\bi{\begin{itemize}}
\def\ei{\end{itemize}}
\def\te{\tilde e}

\def\tu{\tilde u}

\def\tb{\tilde b}

\def\tst{\tilde t}
\def\ttau{\tilde \tau}

\def\tg{\tilde g}

\def\tq{\tilde q}
\def\tw{\widetilde W}
\def\tz{\widetilde Z}

\def\be{\begin{equation}}  
\def\ee{\end{equation}}  

\newcommand\prd[3]{{\it Phys.\ Rev.\ }{\bf D #1} (#2) #3}
\newcommand\prep[3]{{\it Phys.\ Rept.\ }{\bf #1} (#2) #3}
\newcommand\prl[3]{{\it Phys.\ Rev.\ Lett.\ }{\bf #1} (#2) #3}
\newcommand\plb[3]{{\it Phys.\ Lett.\ }{\bf B #1} (#2) #3}
\newcommand\njp[3]{{\it New\ J.\ Phys.\ }{\bf #1} (#2) #3}
\newcommand\jhep[3]{{\it J. High Energy Phys.\ }{\bf #1} (#2) #3}
\newcommand\app[3]{{\it Astropart.\ Phys.\ }{\bf #1} (#2) #3}

\newcommand\ijmpd[3]{{\it Int.\ J.\ Mod.\ Phys.\ }{\bf D #1} (#2) #3}

\newcommand\npb[3]{{\it Nucl.\ Phys.\ }{\bf B #1} (#2) #3}
\newcommand\epjc[3]{{\it Eur.\ Phys.\ J. }{\bf C #1} (#2) #3}

\newcommand\zpc[3]{{\it Z.\ Physik }{\bf C #1} (#2) #3}
\newcommand{\hepph}[1]{hep-ph/#1}

\newcommand{\astroph}[1]{astro-ph/#1}

\begin{document}
\begin{titlepage}

\begin{flushright}
OU-HEP/110605
\end{flushright}

\vspace*{0.5cm}
\begin{center}
{\Large \bf 
Exploring neutralino dark matter resonance annihilation \\
via $bA,bH \to b\mu^+\mu^-$ at the LHC}
\\
\vspace{1.2cm} \renewcommand{\thefootnote}{\fnsymbol{footnote}}
{\large 
Howard Baer$^{1}$\footnote[1]{Email: baer@physics.ou.edu},
Alexander Belyaev$^{2}$\footnote[2]
{Email: a.belyaev@soton.ac.uk},
Chung Kao$^1$\footnote[3]{Email: kao@physics.ou.edu}
and  Patrik Svantesson$^2$\footnote[4]
{Email: p.svantesson@soton.ac.uk}} \\
\vspace{1.2cm} \renewcommand{\thefootnote}{\arabic{footnote}}
{\it 
$^1$
Homer L. Dodge Department of Physics and Astronomy, 
University of Oklahoma, Norman, OK 73019, USA \\
$^2$NExT Institute: School of Physics \& Astronomy, Univ. of Southampton, UK\\
Particle Physics Department, Rutherford Appleton Laboratory, UK
}

\end{center}

\vspace{0.5cm}
\begin{abstract}
One of the main channels which allows for a large rate of
neutralino dark matter annihilation in the early Universe is
via the pseudoscalar Higgs $A$-resonance.  In this case, the
measured dark matter abundance can be obtained in the minimal
supergravity (mSUGRA) model when $\tan\beta\sim 50$ and
$2m_{\tz_1}\sim m_A$. We investigate the reaction $pp\to
b\phi\to b\mu^+\mu^- +X$ (where $\phi =A$ or $H$)  at the CERN
LHC where requiring the tag of a single $b$-jet  allows for
amplification of the signal-to-background ratio. The rare but
observable Higgs decay to muon pairs allows for a precise
measurement of the Higgs boson mass and decay width. We
evaluate signal and background using CalcHEP, with muon energy
smearing according to the CMS detector.  
We find that the Higgs width ($\Gamma_A$) can typically be determined 
with the accuracy  up to $\sim 8\%$ ($\sim 17\%$) for $m_A\sim 400$ (600) GeV 
assuming $10^3$ fb$^{-1}$ of integrated luminosity.  
Therefore, the $pp\to b\phi\to b\mu^+\mu^- +X$ process provides a 
unique possibility for direct $\Gamma_A$  measurement at the LHC.
While the Higgs width is correlated
with the parameter $\tan\beta$ for a  given value of $m_A$,
extracting $\tan\beta$ is complicated by an overlap of the $A$ and 
$H$ peaks, radiative corrections to the  $b$ and $\tau$
Yukawa couplings, and the possibility that SUSY decay modes of
the Higgs may be open. In the case where a dilepton mass edge
from $\tz_2\to\ell^+\ell^-\tz_1$ is visible, it should be
possible to test the relation that $2m_{\tz_1}\sim m_A$.
\noindent 
\vspace{0.8cm}


\end{abstract}


\end{titlepage}

\section{Introduction}
\label{sec:intro}

The lightest neutralino $\tz_1$ of $R$-parity conserving supersymmetric 
(SUSY) models is often touted as an excellent WIMP candidate for 
cold dark matter (CDM) in the universe~\cite{wimp}. 
However, in SUSY models where the $\tz_1$ is mainly bino-like, 
the natural value of the relic density $\Omega_{\tz_1}h^2$ is in 
the 1-100 range~\cite{bbox}, which is far beyond the WMAP 
observation~\cite{wmap7}, 
\be
\Omega_{CDM}h^2\equiv \rho_{CDM}/\rho_c =
\Omega_{CDM}h^2=0.1123\pm 0.0035\ \ \ 68\%\ CL ,
\label{eq:Oh2}
\ee
where $h=0.74\pm 0.03$ is the scaled Hubble constant\cite{href}. 
To gain accord between theory and observation, special neutralino
annihilation mechanisms must be invoked. These include:
({\it i}). co-annihilation (usually involving $\tz_1$ with a
stau\cite{stau}, stop\cite{stop} or chargino\cite{ino}), 
({\it ii}). tempering the neutralino composition\cite{wtn} so it is 
a mixed bino-higgsino (as occurs in the hyperbolic branch/focus point (HB/FP) region of
mSUGRA\cite{hb_fp}) or mixed bino-wino state or 
({\it iii}). annihilation through the light ($h$) or heavy Higgs boson 
resonance $(A\ and/or\ H$)~\cite{Afunnel}. 

In this paper, we are concerned with testing the latter annihilation 
mechanism, which occurs if $2m_{\tz_1}\simeq m_A$. 
The $A$-resonance annihilation mechanism already occurs in the
paradigm minimal supergravity (mSUGRA or CMSSM) model, which
serves as a template for many investigations into SUSY phenomenology.
The mSUGRA parameters at the grand unification (GUT) scale include
\be
m_0,\ m_{1/2},\ A_0,\ \tan\beta\ \ {\rm and}\ \ sign(\mu ),
\ee
where $m_0$ is a common scalar mass, 
$m_{1/2}$ is a common gaugino mass, $A_0$ is a common trilinear term and 
$\tan\beta$ is the ratio of Higgs field vacuum expectation values (VEVs). 
The superpotential Higgs mass term $\mu$ has its magnitude, but not sign, 
determined by radiative breaking of electroweak symmetry (REWSB), 
which is seeded by the large top quark Yukawa coupling.

In mSUGRA, as $\tan\beta$ increases, the $b$- and $\tau$-
Yukawa couplings -- $f_b$ and $f_\tau$ --  also increase, and in
fact their GUT scale values may become comparable to $f_t$ for
$\tan\beta\sim 50$. In this case, the up and down Higgs soft
masses $m_{H_u}^2$ and $m_{H_d}^2$ run  under renormalization
group evolution to nearly similar values at the weak scale.
Since at the weak scale $m_A^2\sim
m_{H_d}^2-m_{H_u}^2$\cite{nuhm},  we find that as $\tan\beta$
increases, the value of $m_A$ decreases\cite{ltanb}, until
finally the condition $2m_{\tz_1}\simeq m_A$ is reached,
whereupon neutralino annihilation through the $A$-resonance
may take place. Another condition that occurs at large
$\tan\beta$ is that since the $b$- and $\tau$ Yukawa couplings
are growing large, the partial widths $\Gamma (A\to b\bar{b})$
and $\Gamma (A\to \tau\bar{\tau})$ also grow,  and the $A$
width becomes very large (typically into the tens of GeV
range).  In this case, a wide range of parameter space
actually accommodates $\tz_1\tz_1$ annihilation through $A,\
H$, and the value of $2m_{\tz_1}$ may be a few partial widths
off resonance since in the relic density  calculation the
$\tz_1\tz_1$ annihilation rate times relative velocity must be
thermally averaged. The question we wish to address here is:
how well may one identify the cosmological scenario of 
{\it neutralino annihilation through the heavy Higgs resonance}
via measurements at the CERN LHC?

Since the $b$-quark Yukawa coupling increases with
$\tan\beta$,  so do the Yukawa-induced Higgs production cross
sections such as  $b\bar{b}\to A$, $bg\to bA$ and
$gg,q\bar{q}\to b\bar{b}A$. The presence of additional high
$p_T$ $b$-jets in the final state for the second and third of
these reactions allows one to tag the $b$-quark related
production mechanisms, and also allows for a cut which rejects
SM backgrounds -- which don't involve the enhanced $b$ Yukawa
coupling -- at low cost to signal. The second of these
reactions, which is tagged by a single $b$-jet in the final
state,  occurs at an order of magnitude greater cross section
than $b\bar{b}A$ production at the LHC\cite{willen}.

The $A$ and $H$ Higgs bosons are expected to dominantly decay
to $b\bar{b}$ and $\tau\bar{\tau}$ final states. Then, the
$bb\bar{b}$ or $b\tau\bar{\tau}$ modes offer a substantial LHC
reach for $A$ and $H$,  especially at large
$\tan\beta$\cite{lhcHiggsreach}.  Along with these decay
modes, the decay $A,\ H\to\mu^+\mu^-$ has been found to be
very useful~\cite{stepanov}.  
Since the $f_\mu$ Yukawa coupling constant also
increases with $\tan\beta$,  this mode maintains its branching
fraction -- typically at the $10^{-4}$ level -- 
even in the face
of increasing $A\to b\bar{b}$ partial width. It also offers
the advantages in that the two high $p_T$ isolated muons are
easy to tag, and the reconstruction of the invariant mass
of muon pair, $m_{\mu^+\mu^-}$, allows a
high precision measurement of the $A$ mass and width,
$\Gamma_A$. In fact, is was shown in Ref. \cite{ddkm} that the
LHC discovery potential for $A\to \mu^+\mu^-$ is greatest in
the $b\mu^+\mu^-$ mode at large $\tan\beta$, compared to
$\mu^+\mu^-$, or $b\bar{b}\mu^+\mu^-$.  We will adopt the
$pp\to b A,\ bH +X$ production mode along with decay to muon
pairs as a key to explore neutralino annihilation via the
$A$-resonance in this paper.

\begin{figure}[htbp]
\begin{center}
\epsfig{file=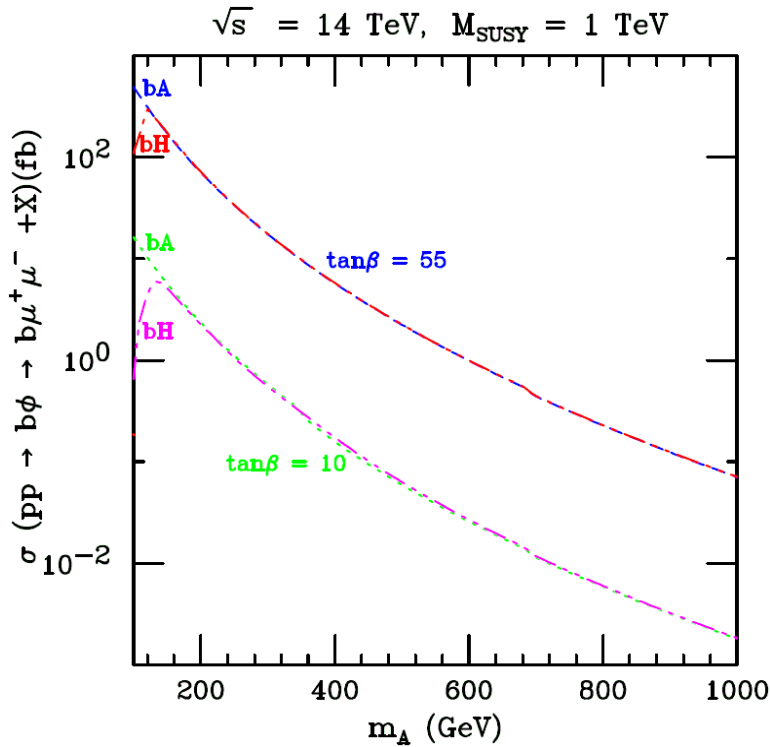,width=0.6\textwidth}
\end{center}
\vspace*{-4mm}
\caption{\it 
The total production cross section for $pp\to b\phi\to b\mu^+\mu^- +X$ 
versus $m_A$ in fb at 
LHC with $\sqrt{s}=14$ TeV. We show results for $\phi =A$ and $H$, and for
$tan\beta =10$ and 55.
}\label{fig:sigbmm} 
\end{figure}

In Fig. \ref{fig:sigbmm}, we show the leading order cross
section for  $pp\to b\phi\to b\mu^+\mu^- +X$ production versus
$m_A$ at LHC with $\sqrt{s}=14$ TeV.  We show curves for $\phi
=A$ and $H$, and for $\tan\beta =10$ and $55$. Several
features are worth noting.
\bi
\item The cross sections for $A$ and $H$ production are nearly identical, 
except at very low $m_A$ values, where substantial mixing between $h$ 
and $H$ occurs.
\item The total production cross section increases by a factor of $\sim 40$ 
in moving from $\tan\beta =10$ to $\tan\beta =55$. This reflects the 
corresponding increase in $b$-quark Yukawa coupling $f_b$, and goes as 
$f_b^2$ in the total production cross section.
\item In spite of the small $A,\ H\to\mu^+\mu^-$ branching fraction of 
$\sim 10^{-4}$, the cross section for $b\mu^+\mu^-$ production via the 
Higgs remains large, varying between over $10^2$ fb for low $m_A$ to 
$\sim 10^{-1}$ fb for $m_A\sim 1$ TeV when $\tan\beta$ is large.
For LHC integrated luminosities ($L$) of order $10^2-10^3$ fb$^{-1}$, 
these rates should be sufficient at least to extract the $A$ and/or $H$ 
mass bump. 
\item 
In addition, the factorization scale 
and the renormalization scale are chosen to be 
$\mu_F = \mu_R = m_\phi/4$ with $\phi = A, H$. 
This choice of scale effectively reproduces the effects of next-to-leading
order (NLO) corrections\cite{willen}.
\ei

The remainder of this paper is organized as follows. 
In Sec.~\ref{sec:msugra}, we review neutralino annihilation via 
the $A$-resonance in the mSUGRA model, and the reach of LHC 
for $A,\ H\to \mu^+\mu^-$ in mSUGRA parameter space for various values of 
integrated luminosities.
In Sec.~\ref{sec:sim}, we present our methods and results from Monte Carlo simulations 
for $A,\ H$ production and decay to muons. 
In Section 4, we  present our strategy to extract Higgs masses ($m_{A,H}$) and 
Higgs widths ($\Gamma_{A,H}$), and show the expected precision 
that LHC might be expected to attain in measuring $m_A$ and $\Gamma_A$.
In Sec.~\ref{sec:conclude} (Conclusions), we comment on how these measurements will help 
ascertain when $A$-resonance annihilation might be the major annihilation 
reaction for neutralino dark matter in the early universe.

\section{The $A$-resonance annihilation region in mSUGRA}
\label{sec:msugra}

\begin{figure}[htbp]
\begin{center}
\epsfig{file=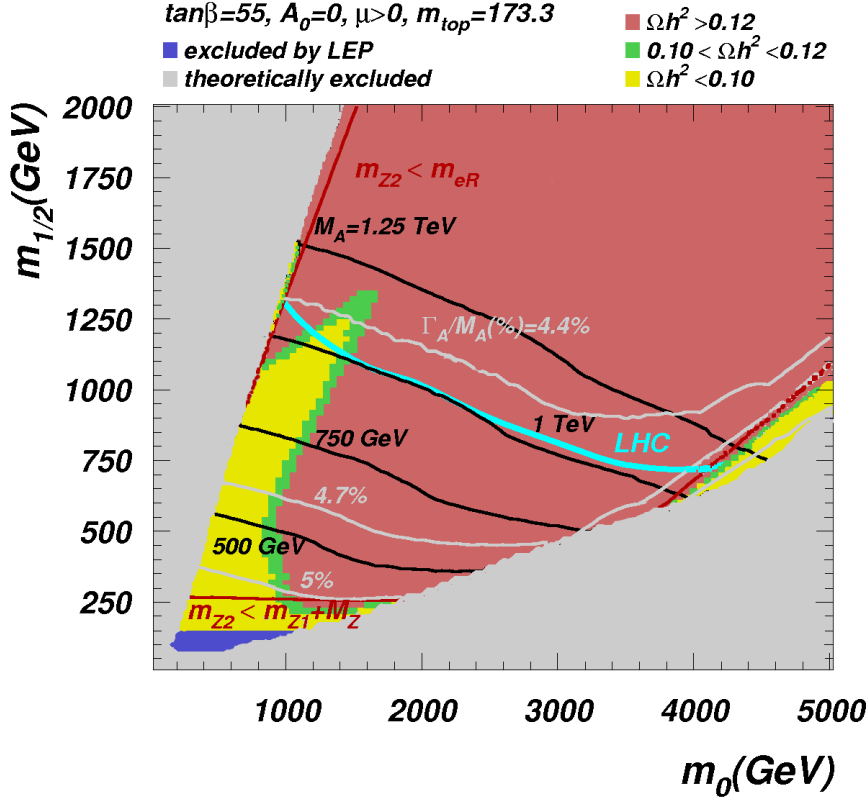,width=0.8\textwidth}
\end{center}
\vspace*{-4mm}
\caption{\it 
The $m_0\ vs.\ m_{1/2}$ frame of the mSUGRA model for $\tan\beta =55$, $A_0=0$,
$\mu >0$ and $m_t=173.3$ GeV. The green and yellow regions provide a thermal
neutralino abundance in accord with
WMAP measurements of the dark matter density. We also show contours
of $m_A$ and $\Gamma_A/m_A$, and also show the LHC reach for SUSY with 100 fb$^{-1}$ and $\sqrt{s}=14$ TeV.
}\label{fig:sug55} 
\end{figure}

In this section, we would like to map out the portions of the $A$-resonance
annihilation parameter space which are potentially accessible to LHC searches.
Figure~\ref{fig:sug55} shows our results in the $(m_0, m_{1/2})$ 
plane of the mSUGRA model for $A_0=0$, $\tan\beta =55$ and $\mu >0$.
The green-shaded region has a relic density\footnote{
The neutralino relic density is computed with the IsaReD\cite{isared} 
subroutine of Isajet 7.80\cite{isajet}.}  of $0.1<\Omega_{\tz_1}h^2<0.12$, 
while the yellow-shaded region has $\Omega_{\tz_1}h^2<0.1$.
The red-shaded region has too large a thermal neutralino abundance 
$\Omega_{\tz_1}h^2>0.12$, and so is excluded under the assumption of 
a standard cosmology with neutralino dark matter.
The gray region is excluded because either REWSB breaks down (right side), 
or we find a stau as the lightest SUSY particle (LSP) (left side). 
The blue shaded region is excluded 
by LEP2 searches for chargino pair production, {\it i.e.} 
$m_{\tw_1}<103.5$ GeV.

The $A$-resonance annihilation region is plainly visible on the plot. 
We also show the SUSY reach of the CERN LHC assuming $\sqrt{s}=14$ TeV 
and 100 fb$^{-1}$ of integrated luminosity, taken from
Fig. 5 of Ref. \cite{bbbkt}. 
The LHC reach is mainly determined by the total cross section for
$\tg\tg$, $\tg\tq$ and $\tq\tq$ production, followed by their subsequent 
cascade decays\cite{cascade} into final states with multi-jets plus 
multi-isolated leptons plus missing transverse energy (MET). 
A hypercube of cuts is examined to extract signal and background rates 
over a variety of cascade decay signal channels. We see that with 
$L = 100$ fb$^{-1}$, LHC can nearly cover the entire $A$-funnel. 
Doubling the integrated luminosity would allow for complete exploration 
of this DM-allowed region. Meanwhile, much of the HB/FP region
is inaccessible to LHC searches, although it should be completely covered by future
WIMP searches by Xe-100 and Xe-1-ton experiments\cite{ofarril,xe100}. 

We also show the contour where $m_{\tz_2}>m_{\te_R}$ and where $m_{\tz_2}<m_{\tz_1}+M_Z$.
In the former region, the decay $\tz_2\to e\te_R\to e^+e^-\tz_1$ will be kinematically open
while in the latter region the 3-body decay
$\tz_2\to\tz_1 e^+e^-$ should be visible. In either case, the dilepton mass edge $m_{\ell^+\ell^- }$ 
should provide information on $m_{\tz_2}$ and $m_{\tz_1}$\cite{mlledge}.

Next, we would like to know how much of the $A$-funnel region is open to heavy Higgs detection
in the $A\to \mu^+\mu^-$ mode. A parton level study has been performed in Ref. \cite{ddkm}
for $m_A$ values up to 600 GeV. 
Here, we wish to extend these results to much higher $m_A$ values.
In Ref. \cite{ddkm}, the maximal reach for the $A,\ H\to\mu^+\mu^-$ mode 
was found to be in the $pp\to b\phi +X$ channel, where $\phi=A$ or $H$.

The study in Ref. \cite{ddkm} evaluated $pp\to b\phi\to b\mu^+\mu^- +X$ 
production against SM backgrounds coming from 
$bg\to b\mu\mu$, from $gg,\ q\bar{q}\to b\bar{b}W^+W^-$ and
from $gb \to bW^+W^-$ (followed by $W\to\mu\nu_\mu$ decay).
They required the presence of
\begin{itemize}
\item two isolated opposite-sign muons with $p_T>20$ GeV and pseudorapidity $|\eta_\mu |<2.5$ ,
\item one tagged $b$-jet, with $p_T(b-jet)>15\ (30)$ GeV and $|\eta_b |<2.5$
and $b$-jet detection efficiency of $\epsilon_b=60\ (50)$\% for low (high) 
integrated luminosity regimes,
\item $MET<20$ GeV (40 GeV), to reduce backgrounds from $t\bar{t}$ production
for   low (high) integrated luminosity regimes.
\end{itemize}
The number of signal and background events within the mass range $m_\phi\pm\Delta M_{\mu^+\mu^-}$
was examined. Here, $\Delta M_{\mu^+\mu^-}\equiv 1.64\left[(\Gamma_\phi/2.36)^2+\sigma_m^2\right]^{1/2}$, 
with $\Gamma_\phi$ equal to the total width of the Higgs boson, and $\sigma_m$ was the muon mass resolution, 
taken to be 2\% of $m_\phi$. 
The signal is considered to be observable if the lower limit 
on the signal plus background is larger than the corresponding upper limit 
on the background with statistical fluctuations
\begin{eqnarray}
L (\sigma_s+\sigma_b) - N\sqrt{ L(\sigma_s+\sigma_b) } \ge 
L \sigma_b +N \sqrt{ L\sigma_b }
\end{eqnarray}
or equivalently, 
\begin{equation}
\sigma_s \ge \frac{N}{L}\left[N+2\sqrt{L\sigma_b}\right] \, .
\end{equation}
Here $L$ is the integrated luminosity, 
$\sigma_s$ is the cross section of the signal,
and $\sigma_b$ is the background cross section.
The parameter $N$ specifies the level or probability of discovery, 
which is taken to be $N = 2.5$ for a 5$\sigma$ signal.
For $\sigma_b \gg \sigma_s$, this requirement becomes similar to 
\begin{eqnarray*}
N_{\rm SS} = \frac{N_s}{\sqrt{N_b}}
 = \frac{L\sigma_s}{\sqrt{L\sigma_b}} \ge 5 \, ,
\end{eqnarray*}
where 
$N_s$ is the signal number of events, 
$N_b$ is the background number of events,
and $N_{\rm SS} =$ the statistical significance, which is 
commonly used in the literature.

Here, at the first stage of our analysis,
we repeat this calculation, although we extend the results to much 
higher values of $m_\phi$ and higher integrated luminosities.
We also have evaluated  additional possible backgrounds 
to make sure that their contributions are either important or negligible.
In our analysis, we use the CTEQ6L set for PDFs\cite{Kretzer:2003it}
and the QCD scale is set equal to $m_A/4$ for signal and $\hat{s}$
for backgrounds.
The following backgrounds have been evaluated with the respective $K$-factors applied
to take into account Higher order corrections:
\begin{itemize}
\item $gg+q\bar{q}\to W^+W^-b\bar{b} \to \mu^+\mu^-\nu\bar{\nu}$ ($K=2$):
this is the dominant background coming mainly from $t\bar{t}$
production and decay,
\item $bg \to W^+W^- b \to \mu^+\mu^-\nu\bar{\nu}$ ($K=1.3$):
this background is typically at least one order of magnitude below
the first one,
\item $bg  \to b\mu^+\mu^-\nu\bar{\nu}$  ($K=1.3$):
this  background is of the same order of magnitude as 
previous one,
\item $b\gamma  \to b\mu^+\mu^-\nu\bar{\nu}$ ($K=1.3$):
this  background is several times lower than the previous
one and can be considered as a subdominant one, contributing to the total background at
the percent level. It was evaluated using the photon distribution function of the proton
available in CalcHEP,
\item $c g   \to c \mu^+\mu^-\nu\bar{\nu}$ ($K=1.3$):
this  background is of the same order as the previous
one and  therefore, again  contributing to the total background at
the percent level. It was evaluated using a mis-tagging probability for $c$-jet
equal to 10\%.
\item $q g   \to q \mu^+\mu^-\nu\bar{\nu}$ ($K=1.3$):
also, this  background is of the same order as the previous
one, contributing to the total background at
the percent level. It was evaluated using mis-tagging probability for light $q$-jet
equal to 1\%.
\end{itemize}

\begin{figure}[htbp]
\begin{center}
\epsfig{file=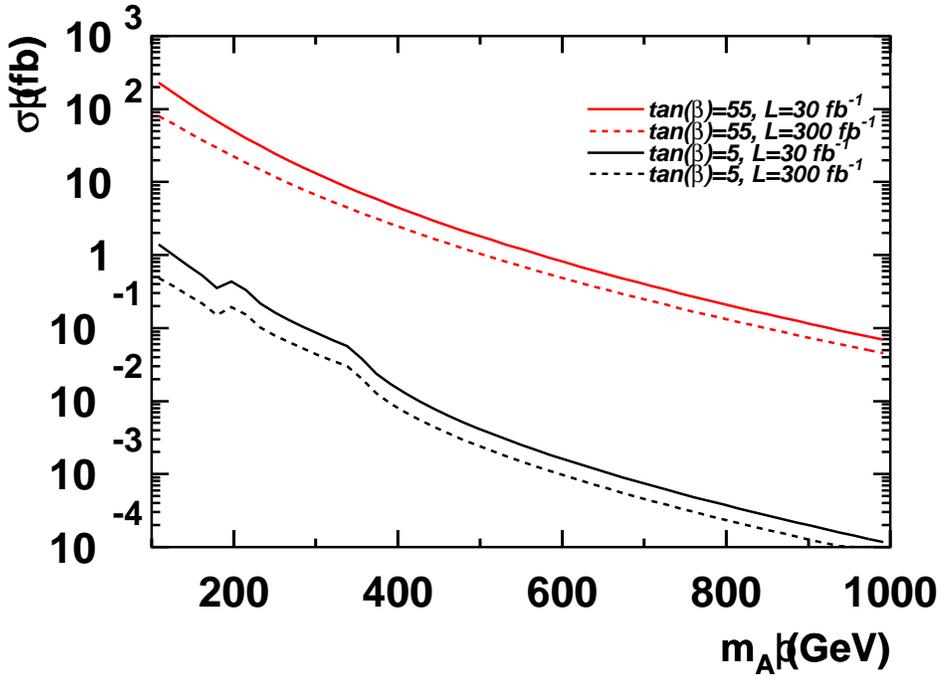,width=0.8\textwidth}
\end{center}
\vspace*{-4mm}
\caption{\it $pp\to \phi b \to \mu^+\mu^- b +X$ signal rates versus $m_A$, 
where $\phi=A,\ H,\ h$ after application of kinematical cuts and efficiency of $b$-tagging
for $\tan\beta=5$ (black lines) and $\tan\beta=55$ (red lines).
Results for low and high luminosity regimes are denoted by solid and dashed lines respectively.
\label{fig:signal}}
\end{figure}

Fig.~\ref{fig:signal} presents $pp\to \phi^0 b \to \mu^+\mu^- b +X$ 
signal rates versus $m_A$, where $\phi^0=A,H,h$ after application of 
kinematical cuts and efficiency of $b$-tagging for $\tan\beta = 5$ (black lines)
and $\tan\beta =55$ (red lines).
Results for low and high luminosity regimes are denoted by solid and dashed lines respectively.

\begin{figure}[htbp]
\begin{center}
\epsfig{file=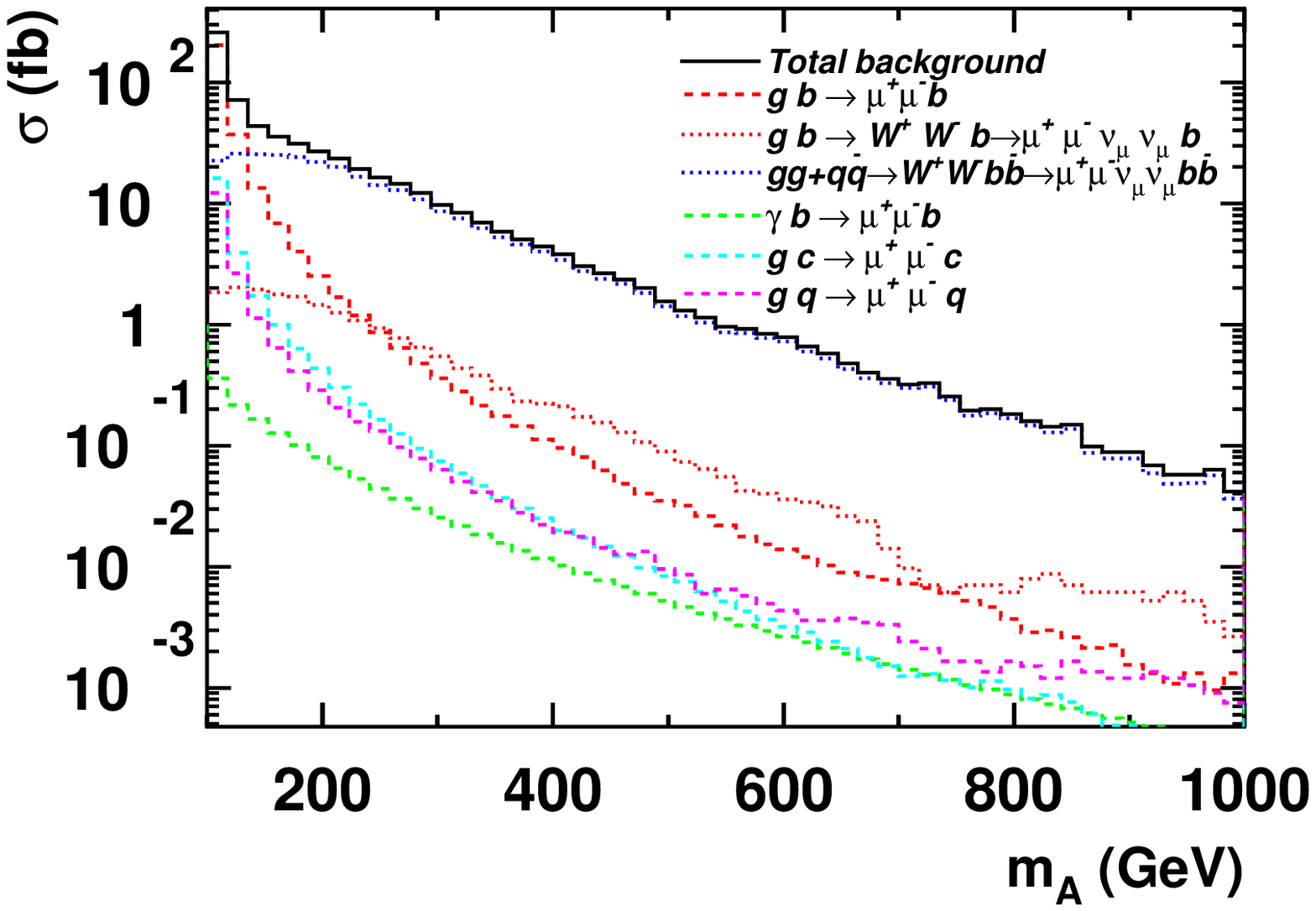,width=0.5\textwidth}%
\epsfig{file=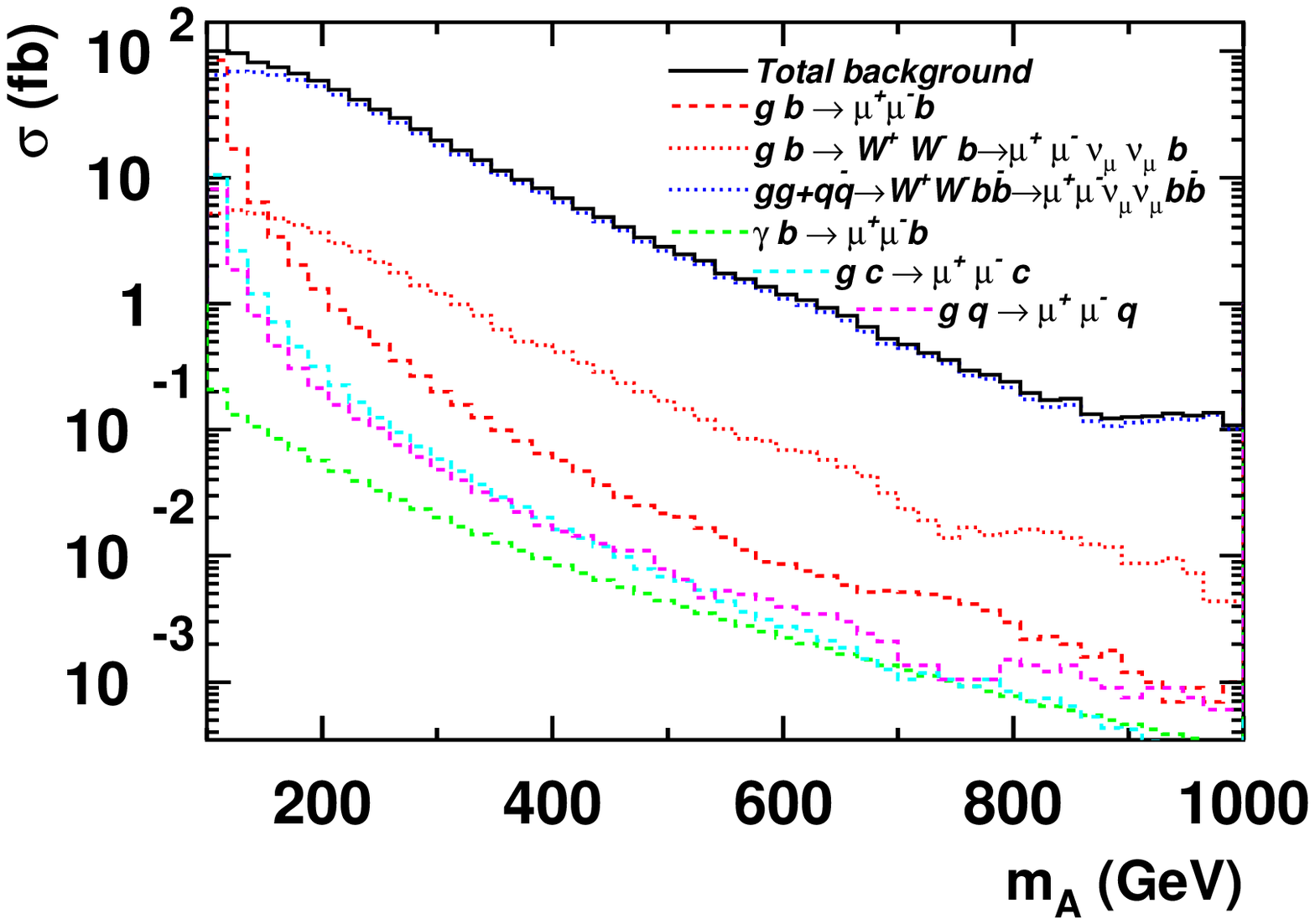,width=0.5\textwidth}
\end{center}
\vspace*{-4mm}
\caption{\it
Rates for various backgrounds for $\mu^+\mu^- b$ signature versus $m_A$
after application of kinematical cuts and efficiency of $b$-tagging
for $\tan\beta= 30$.
Results for low and high luminosity regimes are 
presented  in left and right  frames respectively.
\label{fig:background}} 
\end{figure}

In Fig.~\ref{fig:background}, we present rates 
for various backgrounds described above for $\mu^+\mu^- b$ signature versus $m_A$
after application of kinematical cuts and efficiency of $b$-tagging
for an intermediate value of $\tan\beta =30$.
Results for low and high luminosity regimes are presented  in left and right frames respectively.
One can see that indeed the contributions from the last three subdominant backgrounds 
discussed above are at the percent level.

\begin{figure}[htbp]
\begin{center}
\epsfig{file=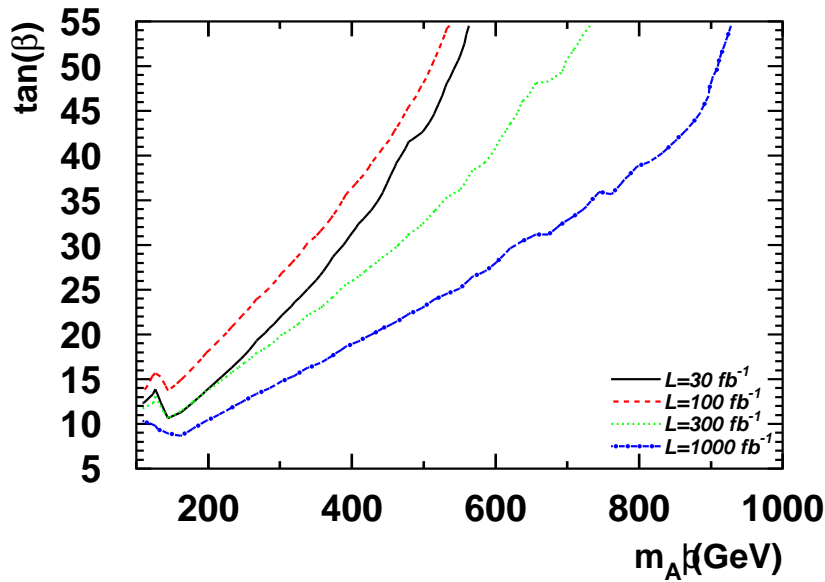,width=0.7\textwidth}
\end{center}
\vspace*{-4mm}
\caption{\it 
LHC reach for the pseudoscalar Higgs $A$ via 
$pp\to bA\to b\mu^+\mu^- +X$ in the $m_A\ vs.\ \tan\beta$ plane
for various possible values of integrated luminosity.
The 30 fb$^{-1}$ reach exceeds the 100 fb$^{-1}$ reach because we use harder cuts in the high
luminosity case.
}\label{fig:reach} 
\end{figure}

Using  signal and background rates from these calculations, we derive 
the LHC discovery reach. The results are shown in Fig. \ref{fig:reach}. 
One should notice an important effect of the cuts 
for low and high luminosity regimes.
The main effect for  LHC reach comes from the $MET$ cut. 
We require $MET<20$ for low luminosity which should leave
signal intact assuming that instrumental missing transverse momentum 
is under control above 20 GeV in the low luminosity regime.
This cut significantly suppresses the leading $t\bar{t}$ background.
For  high luminosity regime we apply $MET<40$
which does not affect signal but significantly increase background.
The overall effect of high luminosity cuts is an increase of the background and decrease of the
signal. Therefore, the discovery potential of the LHC at $L=100 \mbox{ fb}^{-1}$
is slightly lower than at $L=30 \mbox{ fb}^{-1}$. But in the region
sufficiently above the border-line for  LHC discovery potential shown in Fig. \ref{fig:reach},
say for  $m_A=400$~GeV and $\tan\beta=55$, LHC at  $L=100 \mbox{ fb}^{-1}$
provides better statistics and significance as compared to
the $L=30 \mbox{ fb}^{-1}$ case as we show below.
 
Here, we see that for $L=100$ fb$^{-1}$, 
the reach for $b\phi\to b\mu^+\mu^-$ at $\tan\beta\sim 55$ extends to $m_A\simeq 550$ GeV.
For $L=300$ fb$^{-1}$, the reach extends to $m_A\simeq 730$ GeV, and for $L=1000$ fb$^{-1}$, 
the reach extends to $\sim 925$ GeV.

From the results of Fig. \ref{fig:reach}, we can now compare against Fig. \ref{fig:sug55}
to see how much of the $A$-funnel can be explored via the $A,\ H\to\mu^+\mu^-$ decay mode.
To illustrate, we show contours of $m_A=500,\ 750,\ 1000$ and 1250 in Fig. \ref{fig:sug55}.
Thus, for 100 fb$^{-1}$ of integrated luminosity, we expect LHC to be sensitive to 
a $A,\ H\to\mu^+\mu^-$ bump for about half of the $A$-funnel. An integrated luminosity of
300 fb$^{-1}$ covers about three-quarters of the $A$-funnel, while well over $1000$ fb$^{-1}$
will be needed to cover the entire funnel region.

In addition to measuring the value of $m_{A,H}$ via a dimuon mass bump, one may be able to
extract information on the $A,\ H$ widths from the dimuon channel, if LHC experiments have
sufficiently good muon energy reconstruction. To illustrate the values of $\Gamma_A$ that are
expected, we plot contours of $\Gamma_A/m_A$ in Fig. \ref{fig:sug55}.
These range from about 5\% for low $m_{1/2}\sim 250$ GeV, corresponding to $\Gamma_A\sim 15-20$ GeV,
to about 4.4\% for $m_{1/2}\sim 1200$ GeV, where $\Gamma_A\sim 50$ GeV.
\begin{figure}[htbp]
\begin{center}
\epsfig{file=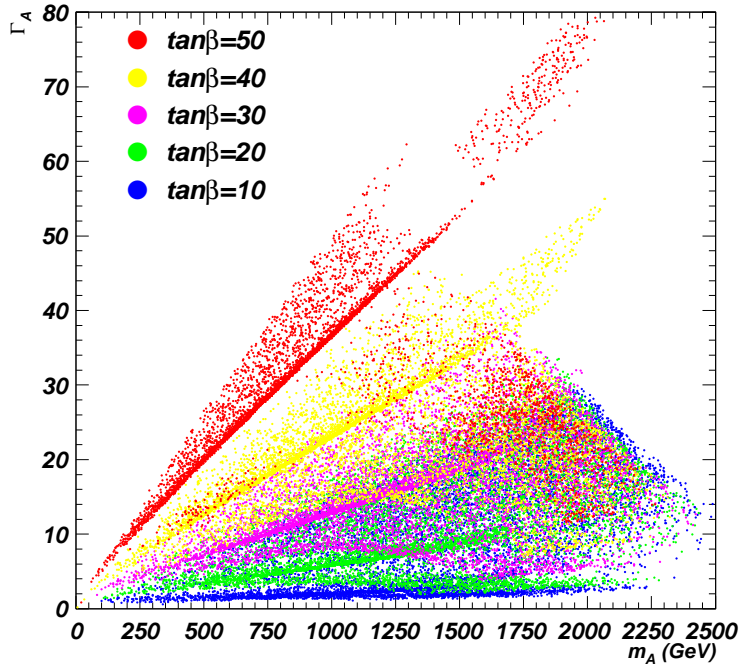,width=0.6\textwidth}
\end{center}
\vspace*{-4mm}
\caption{\it 
Plot of $\Gamma_A\ vs.\ m_A$ from a scan over mSUGRA model parameters for
$\tan\beta =10,\ 20,\ 30,\ 40$ and $50$. 
}\label{fig:GAvsmA} 
\end{figure}
To better illustrate the range of Higgs widths expected in mSUGRA, 
we show in Fig. \ref{fig:GAvsmA} the value of $\Gamma_A$ versus 
$m_A$ after a scan over mSUGRA parameter space for various fixed values 
of $\tan\beta$. Here, we see that indeed as $\tan\beta$ grows, so too 
does $\Gamma_A$. In fact, for a measured value of $m_A$, 
a measurement of $\Gamma_A$ will indicate a rather small window 
of allowed $\tan\beta$ values. 
Naively, one might expect a one-to-one correspondence between $\tan\beta$ 
and $\Gamma_A$ for fixed $m_A$. However, two effects that spread out the 
correlation include:
(i). weak scale threshold corrections to $f_b$ that are large at large 
$\tan\beta$, and depend on the entire SUSY spectrum via loop 
effects~\cite{pbmz}, and 
(ii). various additional SUSY decay modes of the $A$ and $H$ may 
open up~\cite{HtoSUSY}, depending on sparticle masses and mixings.
For instance, if $m_A>2m_{\tw_1}$, then the decay mode 
$A\to\tw_1^+\tw_1^-$ opens up and contributes to the $A$ width. 
Thus, models with lighter SUSY particles should correspond to larger 
$\Gamma_A$ values for a given $m_A$ and $\tan\beta$ value, 
whereas if all non-standard decay modes are closed, 
then the lower range of $\Gamma_A$ that is shown may be expected to occur.
The loop corrections to $f_b$ tend to enhance $f_b$ for $\mu >0$ and diminish
$f_b$ for $\mu <0$, leading to somewhat separated bands for each $\tan\beta$ value.

\section{Detailed simulations for $pp\to bA,\ bH +X$}
\label{sec:sim}

In this section, we present a detailed Monte Carlo study of detection of
$b\phi\to b\mu^+\mu^-$ for a particular case study. The benchmark point we adopt
is known as LCC4 in the study by Battaglia {\it et al.}, Ref. \cite{peskin}.
Some of the mSUGRA parameters and sparticle masses as generated by 
Isajet 7.80 are given in Table \ref{tab:bm}. We use a value of 
$m_t=175$ GeV instead of 178 GeV as in Ref. \cite{peskin} since
the latest Isasugra/IsaReD code gives a relic density of $\Omega_{\tz_1}h^2=0.1$
for the 175 GeV value, and 0.16 for the 178 GeV value.
We also examine later how well the value of $\Gamma_A$ can be measured for benchmark point
BM600 with $m_A=608$ GeV.
%
\begin{table}[htb]\centering
\begin{tabular}{lcc}
\hline
parameter & LCC4 & BM600 \\
\hline
$m_0$      & 380 & 900 \\
$m_{1/2}$  & 420 & 650 \\
$A_0$      & 0 & 0 \\
$\tan\beta$  & 53 & 55 \\
\hline
$\mu$       & 528.2 & 750.7 \\
$m_{\tg}$   & 991.5 & 1502.6 \\
$m_{\tu_L}$ & 973.0 & 1609.0 \\
$m_{\tst_1}$& 713.4 & 1167.9 \\
$m_{\tb_1}$ & 798.9 & 1309.5 \\
$m_{\te_L}$ & 475.4 & 998.2 \\
$m_{\te_R}$ & 412.5 & 931.5 \\
$m_{\ttau_1}$ & 206.6 & 541.7 \\
$m_{\tw_1}$ & 325.7 & 520.1 \\
$m_{\tz_2}$ & 325.4 & 519.5 \\ 
$m_{\tz_1}$ & 172.5 & 274.7 \\ 
$m_A$       & 420.7 & 607.9 \\
$m_H$       & 423.5 & 612.0 \\
$m_h$       & 115.1 & 117.1 \\ \hline
$\Delta a_\mu$ & $35\times 10^{-10}$ & $11\times 10^{-10}$ \\
$BF(b\to s\gamma )$ & $1.9\times 10^{-4}$ & $2.8\times 10^{-4}$ \\
$BF(B_s\to\mu^+\mu^- )$ & $2.8\times 10^{-8}$ & $1.1\times 10^{-8}$ \\
$\Omega h^2_{\tz_1}$ & 0.096 & 0.089 \\
$\sigma (\tz_1 p)$ pb & $1.1\times 10^{-8}$ & $1.7\times 10^{9}$ \\
$\Gamma_A$ & $19.1\ {\rm GeV}$ & $31.9\ {\rm GeV}$ \\
$\Gamma_H$ & $19.2\ {\rm GeV}$ & $32.1\ {\rm GeV}$ \\
\hline
\end{tabular}
\caption{Masses and parameters in~GeV units
for Benchmark points LCC4 (with $m_t=175$ GeV) 
and BM600 (with $m_t=173.3$ GeV) using Isajet 7.80.
}
\label{tab:bm}
\end{table}
\begin{figure}[htbp]
\begin{center}
\epsfig{file=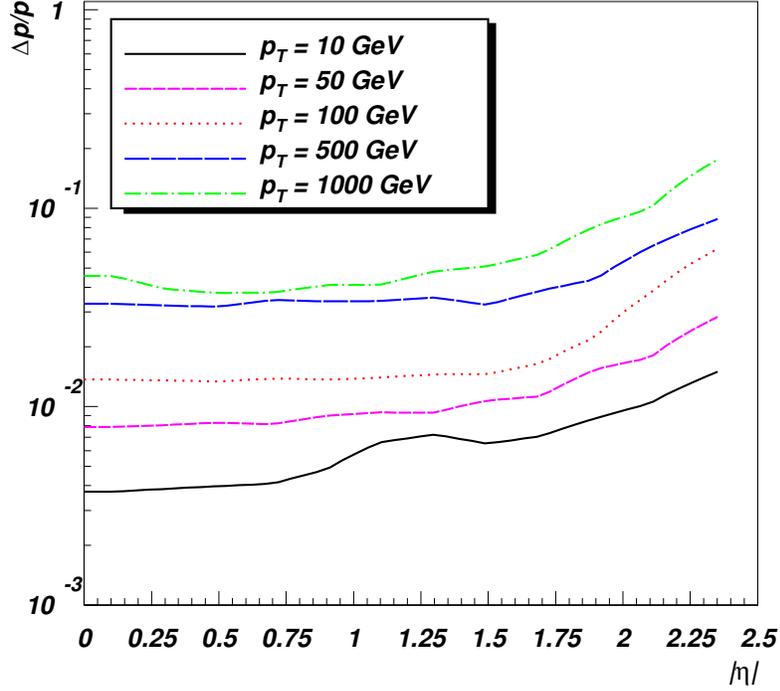,width=0.6\textwidth}
\end{center}
\vspace*{-4mm}
\caption{\it 
Plot of CMS muon smearing function versus $|\eta (\mu )|$ 
for various muon $P_T$ values.
}\label{fig:smearm} 
\end{figure}
\begin{figure}[htbp]
\begin{center}
\epsfig{file=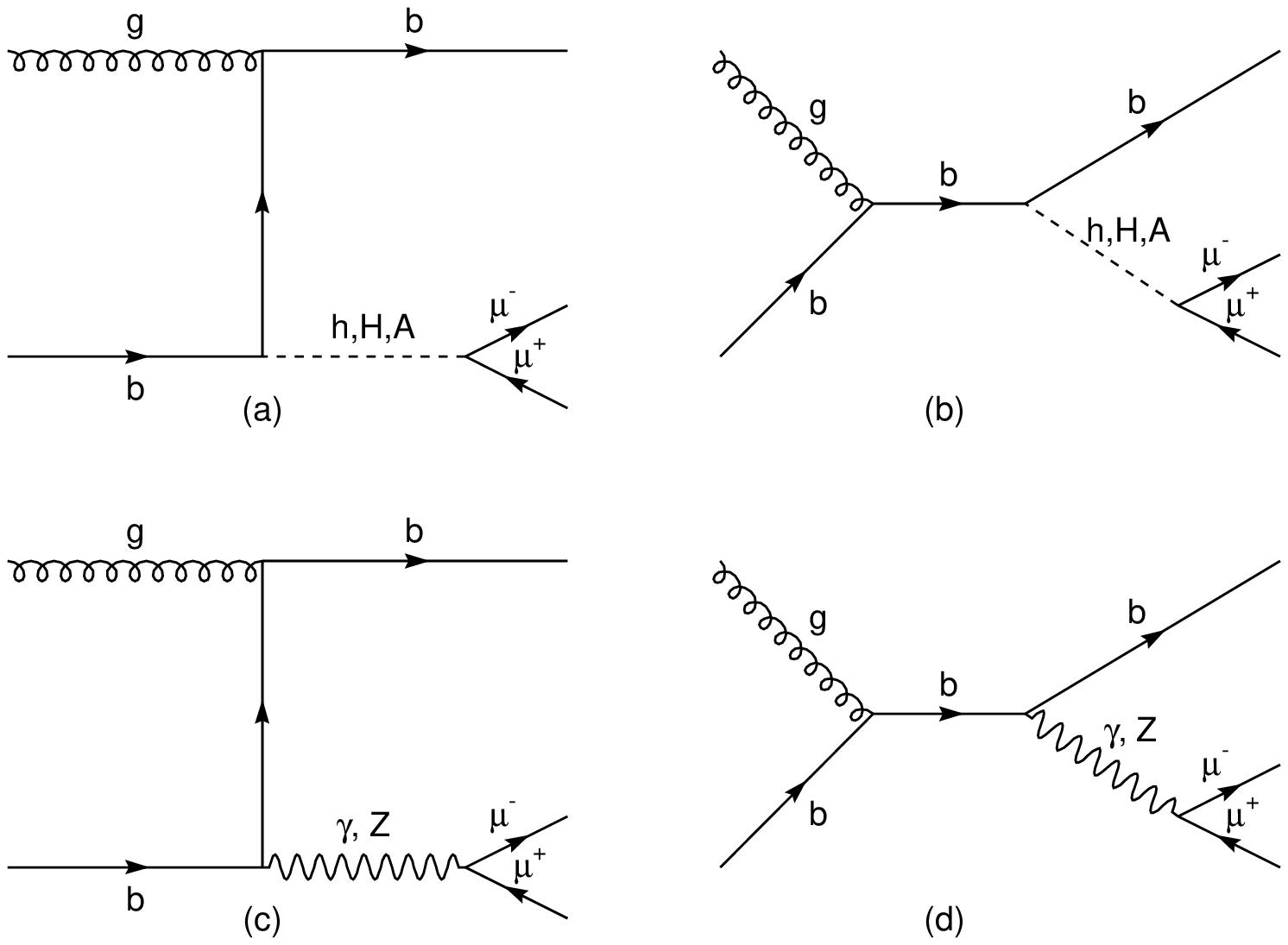,width=0.6\textwidth}
\end{center}
\vspace*{-4mm}
\caption{\it 
Feynman diagrams for $bg \to b\mu^+\mu^-$ in the MSSM.
}\label{fig:diags} 
\end{figure}

The resolution of the dimuon invariant mass, and hence an accurate
measurement of $m_A$ and especially $\Gamma_A$, depends on the
LHC detector's ability to measure the muon's momentum. The muon 
momentum is measured from its amount of bending in the magnetic field
of the detector. Thus, for low energy muon, with a highly curved track,
the muon $\vec{p}$ measurement should be more precise than for 
high energy muons, which have very little track curvature. 
For our studies, we use a CMS muon smearing subroutine, where the
smearing as a function of $|\eta (\mu )|$ is displayed in
Fig. \ref{fig:smearm}, for several muon $p_T$ 
values~\cite{Bayatian:2006zz,Ball:2007zza}.

We begin our MC simulation by calculating $bg\to b\mu^+\mu^-$
production for $pp$ collisions at $\sqrt{s}=14$ TeV 
using CalcHEP\cite{calchep}.
The relevant Feynman diagrams are displayed in Fig. \ref{fig:diags}.
They include not only $A$ and $H$ production and decay, but also
background contributions from $\gamma^*$, $Z^*$ and $h$ production.
\begin{figure}[htbp]
\begin{center}
\epsfig{file=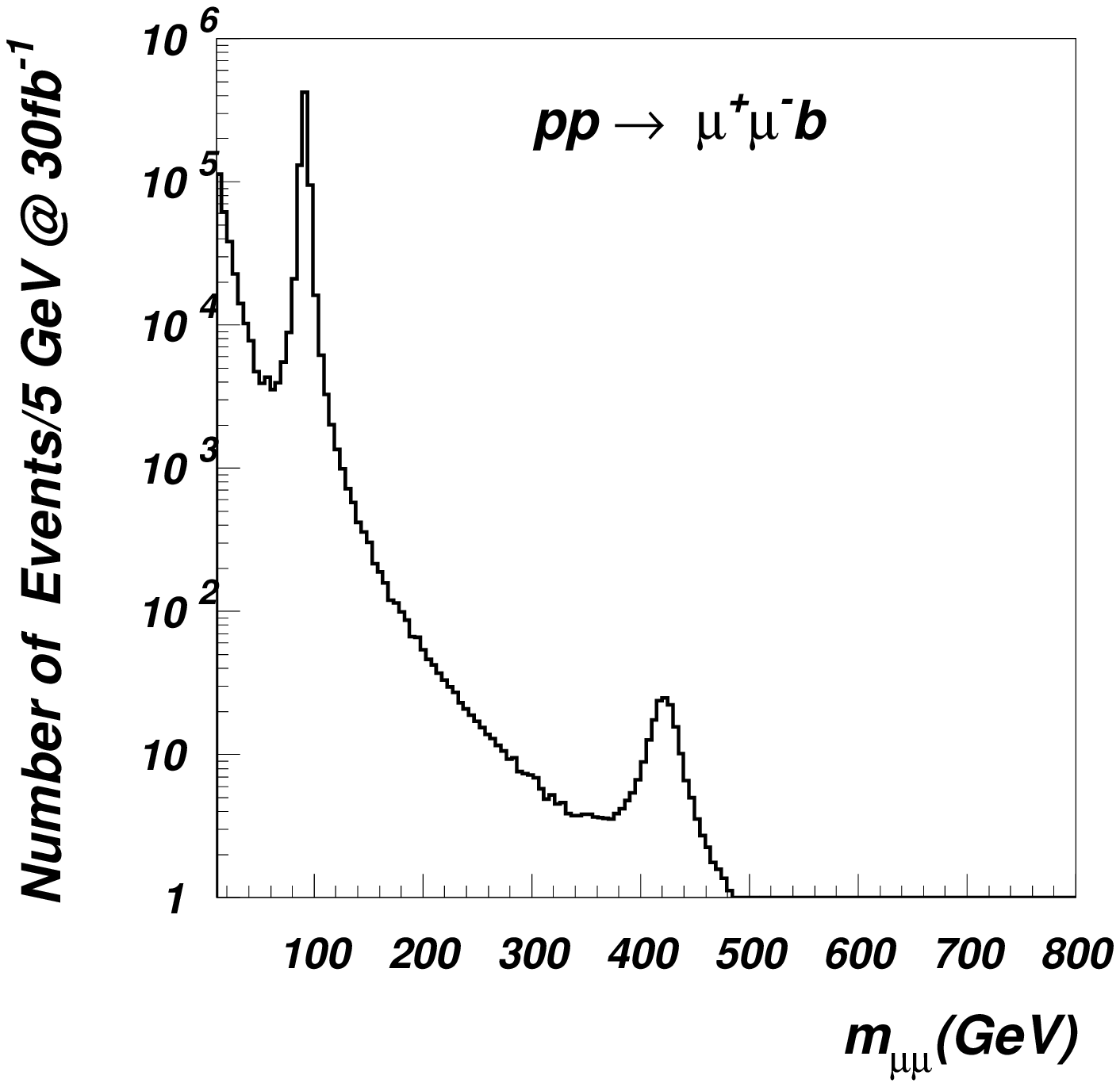,width=0.5\textwidth}
\end{center}
\vspace*{-4mm}
\caption{\it 
Plot of invariant mass distribution of muon pairs $m_{\mu^+\mu^- }$ 
from a CalcHEP MC computation using benchmark LCC4.
}\label{fig:mmm} 
\end{figure}
\begin{figure}[htbp]
\begin{center}
\epsfig{file=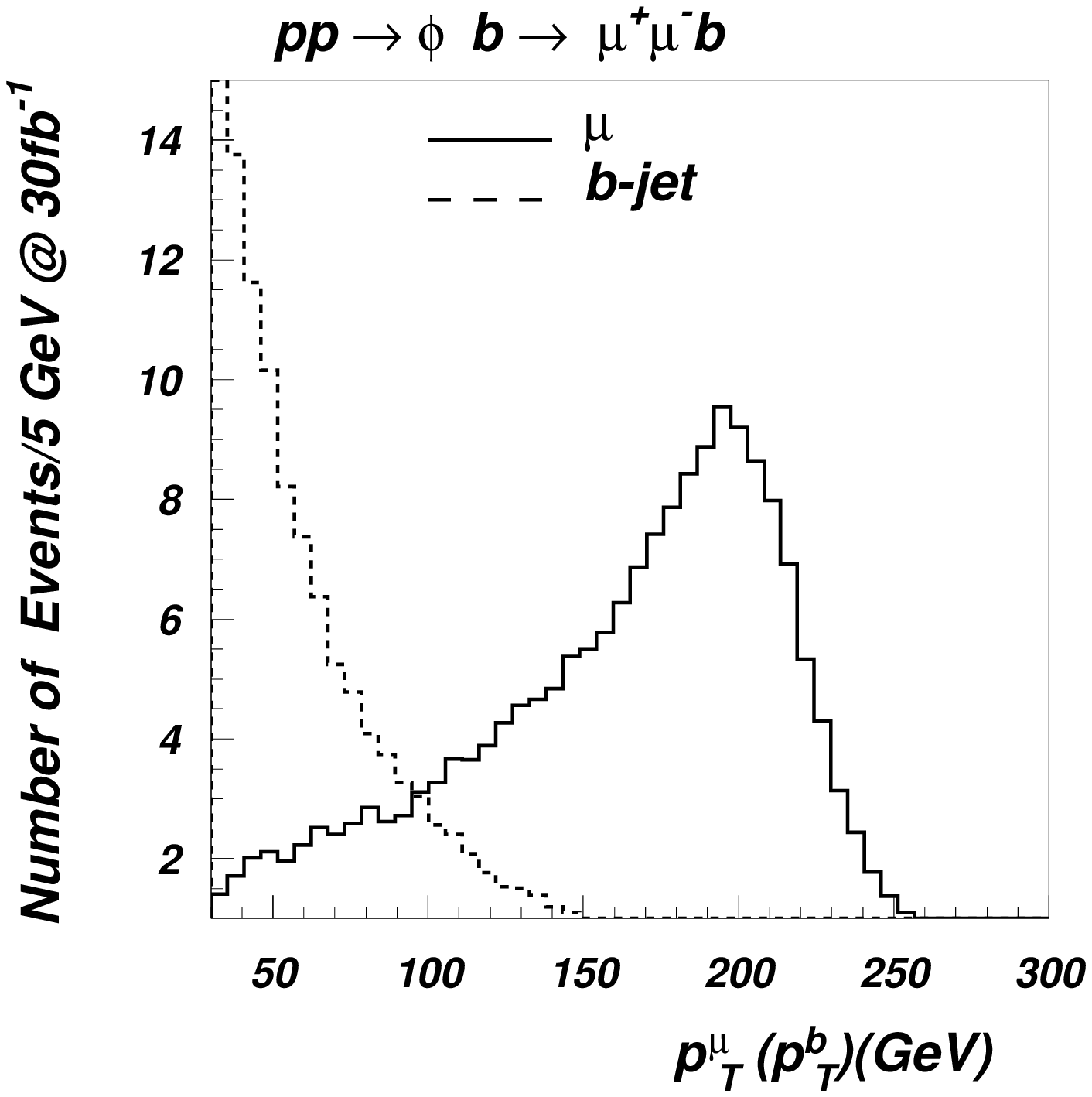,width=0.5\textwidth}
\end{center}
\vspace*{-4mm}
\caption{\it 
Plot of distribution in $p_T(\mu )$ 
(solid line) and  $p_T(b)$ (dashed line) from a CalcHEP MC computation using
benchmark LCC4.
}\label{fig:pt} 
\end{figure}

In Fig. \ref{fig:mmm}, we plot the invariant mass distribution of 
muon pairs $m_{\mu^+\mu^-}$ for $L = 30$ fb$^{-1}$.
For all distributions now and hereafter, we take into account 
detector effects of muon momenta resolution
according to Fig.~\ref{fig:smearm}
using Gaussian smearing applied 
to the particle's momentum generated by CalcHEP at the parton level.
What is clear from the plot is that the $\gamma ,\ Z\to\mu^+\mu^-$
peaks stand out; but also the $A,\ H\to\mu^+\mu^-$ overlapping peak stands out 
well above background levels at $m_A\sim 420$ GeV.

In Fig. \ref{fig:pt}, we plot the muon $p_T$ distribution (solid line)
and  b-jet  $p_T$ distribution (dashed line) 
from $pp\to bA\to b\mu^+\mu^- +X$ production for the LCC4 benchmark. 
The muon $p_T$ distribution peaks at around $p_T\sim m_A/2$, 
but with substantial smearing to either side due to the momentum of
the $A$.
Since the $b$-jets are emitted preferentially in the forward
direction, the $p_T(b)$ distribution peaks at low values, with some smearing
out to values over a hundred GeV.
\begin{figure}[htbp]
\begin{center}
\epsfig{file=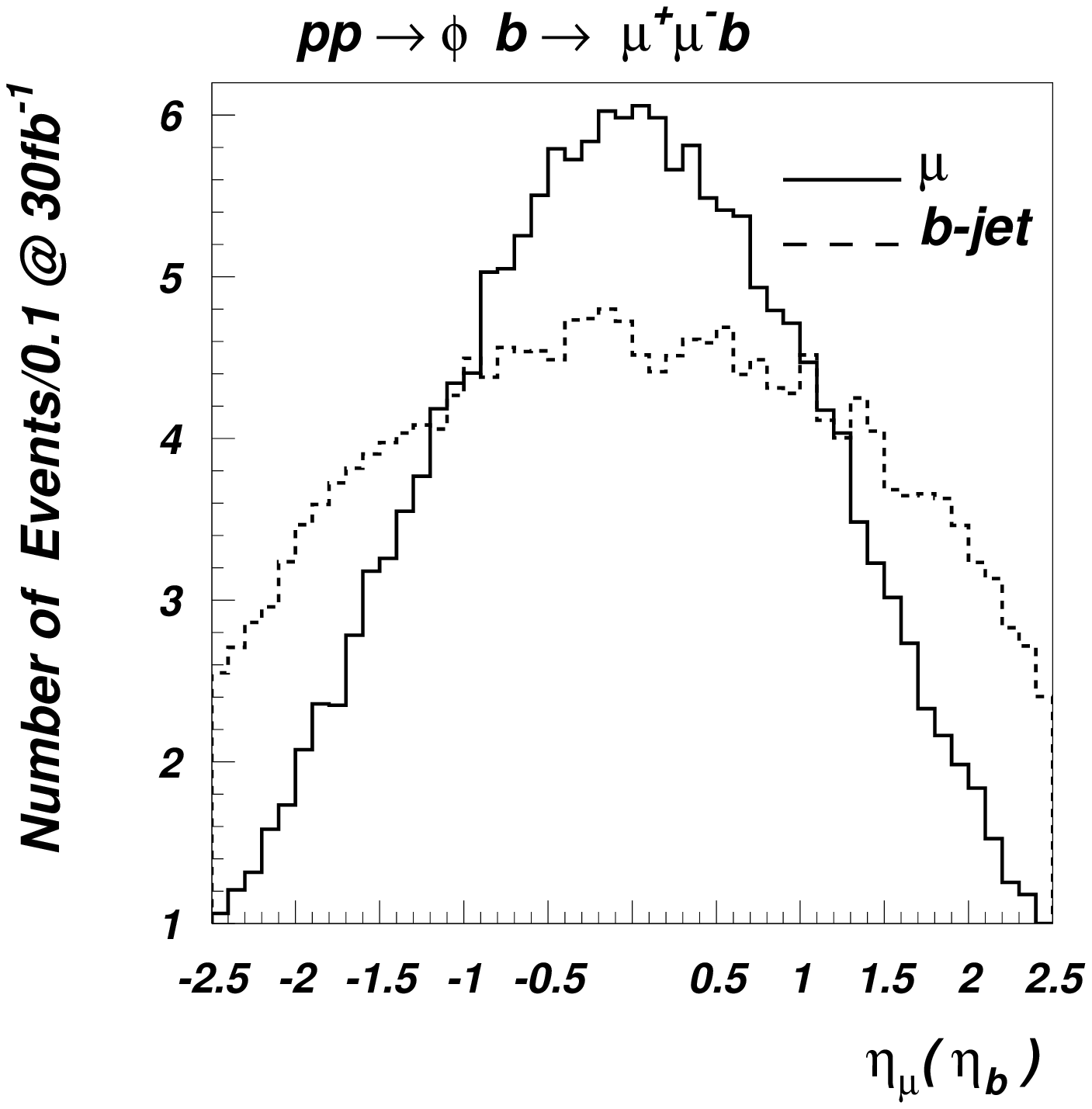,width=0.5\textwidth}
\end{center}
\vspace*{-4mm}
\caption{\it 
Plot of distribution in $\eta(\mu )$ 
(solid line) and  $\eta(b)$ (dashed line) from a CalcHEP MC computation using
benchmark LCC4.
}\label{fig:eta} 
\end{figure}
In Fig. \ref{fig:eta}, we plot the muon (solid line) and $b$-jet (dashed line) 
pseudo-rapidity distributions 
from $pp\to bA\to b\mu^+\mu^- +X$ for the LCC4 benchmark. 
The muon $\eta$ distribution is clearly more central,
while $\eta (b)$ is less central due to its role as an element in QCD
initial state radiation.

\section{Extracting $m_{A,H}$ and $\Gamma_{A,H}$}
\label{sec:mA}

\begin{figure}[htbp]
\begin{center}
\epsfig{file=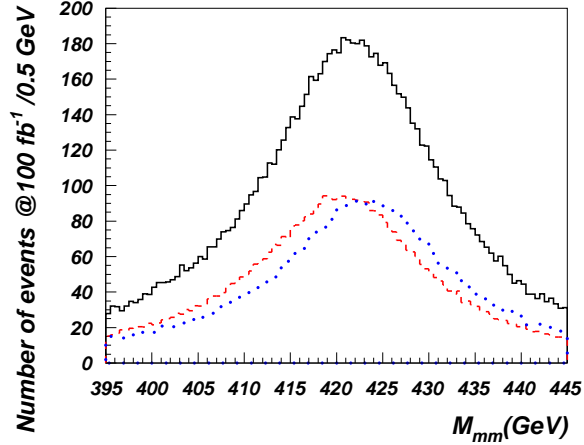,width=0.5\textwidth}
\end{center}
\vspace*{-4mm}
\caption{\it 
Plot of dimuon invariant mass ($M_{\mu^+\mu^-}$) from $bA$ production (red), 
$bH$ production (blue) and sum (black) for benchmark LCC4 with no smearing.
}\label{fig:GAH} 
\end{figure}

Once a dimuon mass bump has been established, the next step is
to fit the invariant mass distribution with a curve which
depends on the Higgs mass and width. A complication occurs
because in our case the $A$ and $H$ masses are only separated by
$\sim 3$ GeV, and so the two peaks are highly overlapping, and essentially
indistinguishable. To see what this means for an ideal measurement, we plot 
in Fig. \ref{fig:GAH} for LCC4 the dimuon invariant mass from just 
the reaction $pp\to bA\to b\mu^+\mu^- +X$ (red curve), and also 
the distribution from $pp\to bH\to b\mu^+\mu^- +X$ (blue curve), 
along with the sum (black curve). 
A direct measurement of these idealized 
distributions of full-width-at-half-max shows indeed that 
$\Gamma_A\simeq 430-410 =20$ GeV, while $\Gamma_H\simeq 433-413 =20$ GeV. 
A measure of the summed distributions provides 
$\Gamma_{A,H}\simeq 433-410=23$ GeV, {\it i.e.} the idealized width 
expectation expanded by the $A,\ H$ mass splitting.
We fit the dimuon invariant mass distribution from all diagrams of 
Fig. \ref{fig:diags} along with muon smearing with the following function 
$F$ of dimuon mass $m$ and 6 fitting parameters
$\Gamma, M, N, \sigma, N_{p1},N_{p2}$:
\begin{equation}
F(m; \Gamma, M, N, \sigma, N_{p1},N_{p2})=
N\int{ B(m',\Gamma, M) 
\times  G(m', m,\sigma) dm'} + N_{p1}\exp(-N_{p2} m),
\end{equation}
where 
$N$ is just a normalization parameter, 
$$B(m',\Gamma, M)=\frac{2}{\pi}\frac{\Gamma^2 M^2}{(m'^2-M^2)^2+m'^4(\Gamma^2/M^2)},$$
$$G(m',m, \sigma)=\frac{1}{\sqrt{2}\pi\sigma}
\exp\left[-\frac{(m'-m)^2}{2\sigma^2}\right].$$
One can see that $F(m; \Gamma, M, N, \sigma, N_{p1},N_{p2})$
is a convolution of the Breit-Wigner resonance function along with Gaussian
detector smearing plus an exponentially dropping 
function describing the background shape.

\begin{figure}[htbp]
\epsfig{file=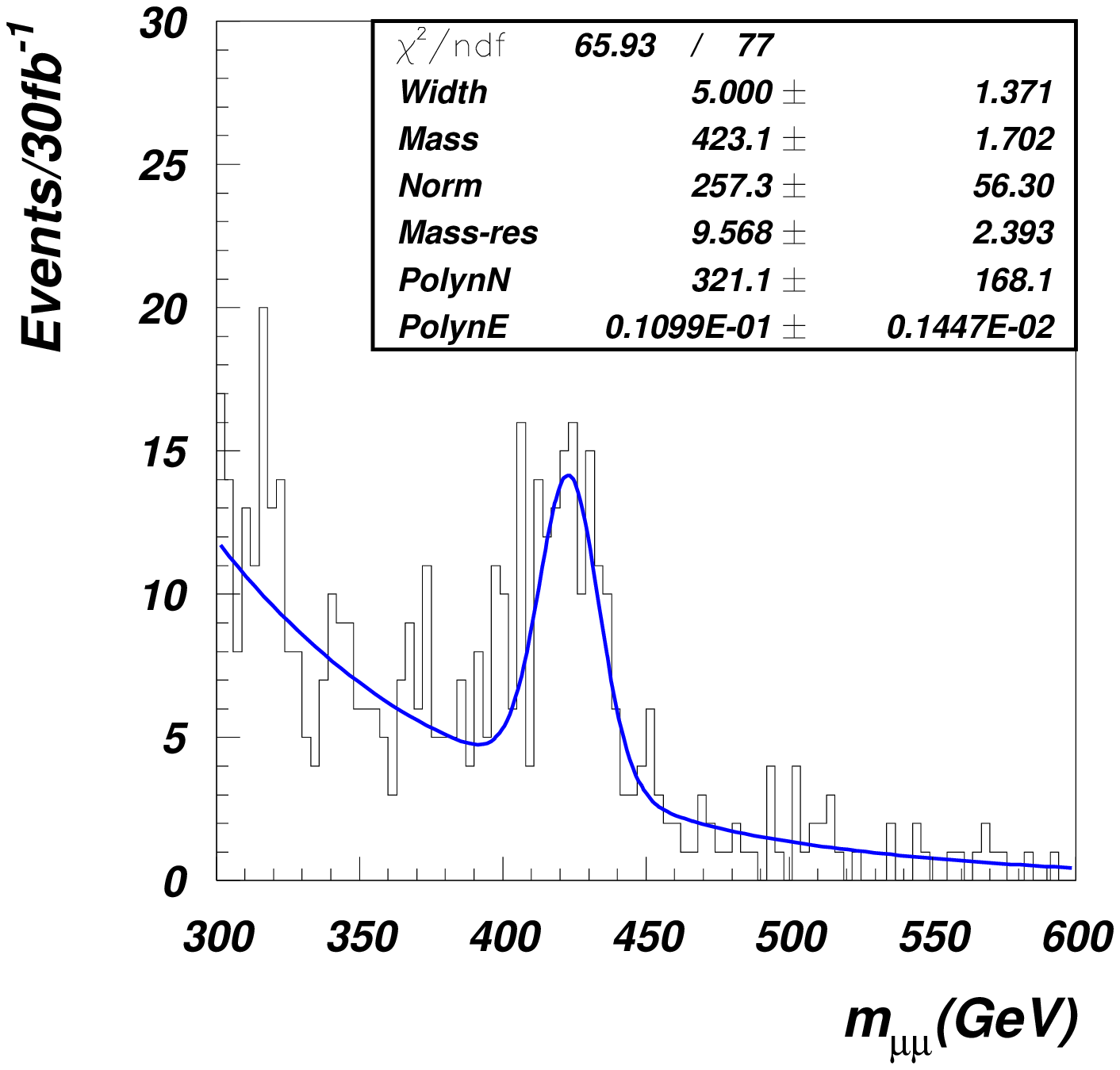,width=0.5\textwidth,height=0.25\textheight}%
\epsfig{file=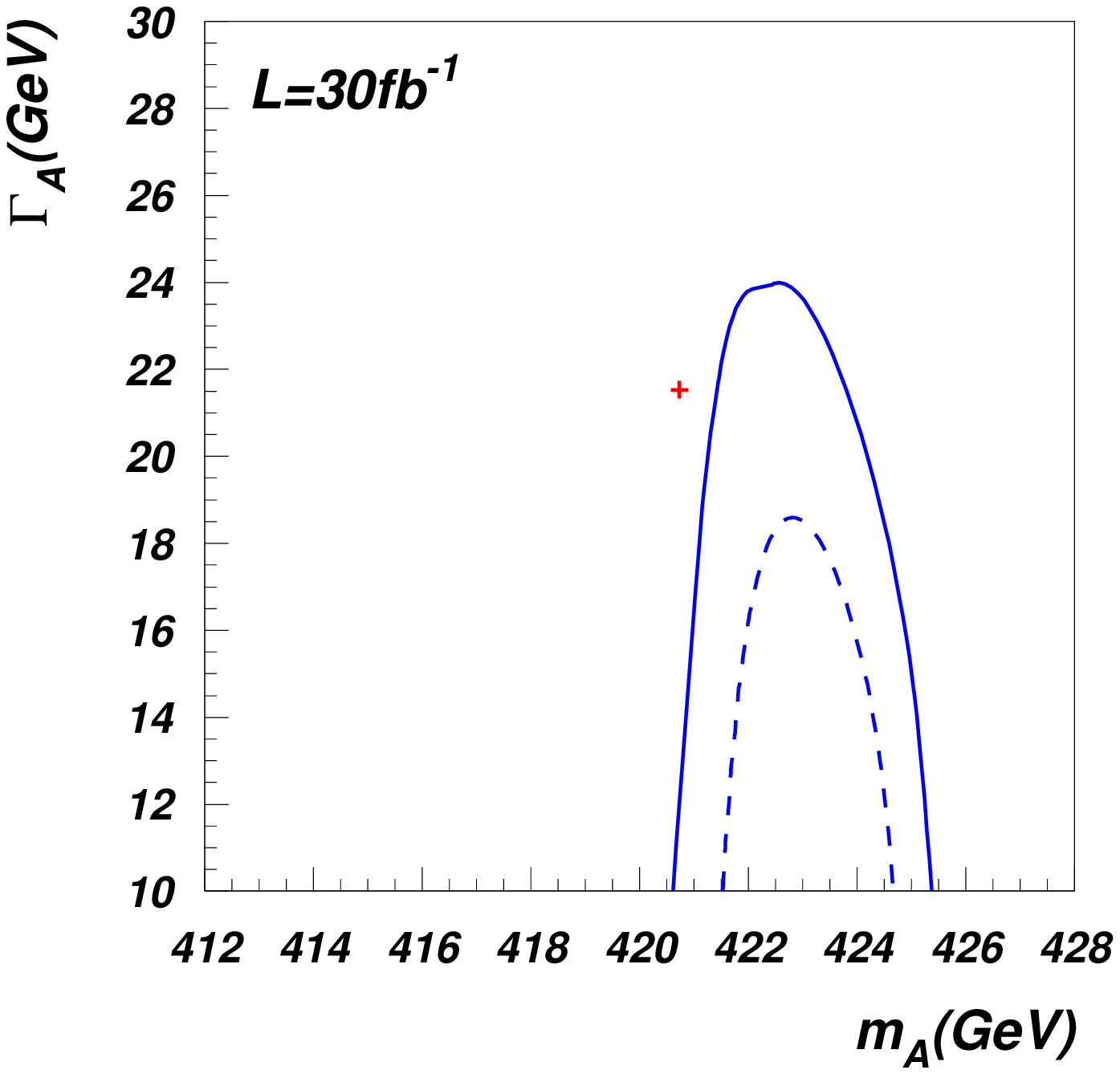,width=0.5\textwidth,height=0.25\textheight}\\
\vskip -1.0cm
\epsfig{file=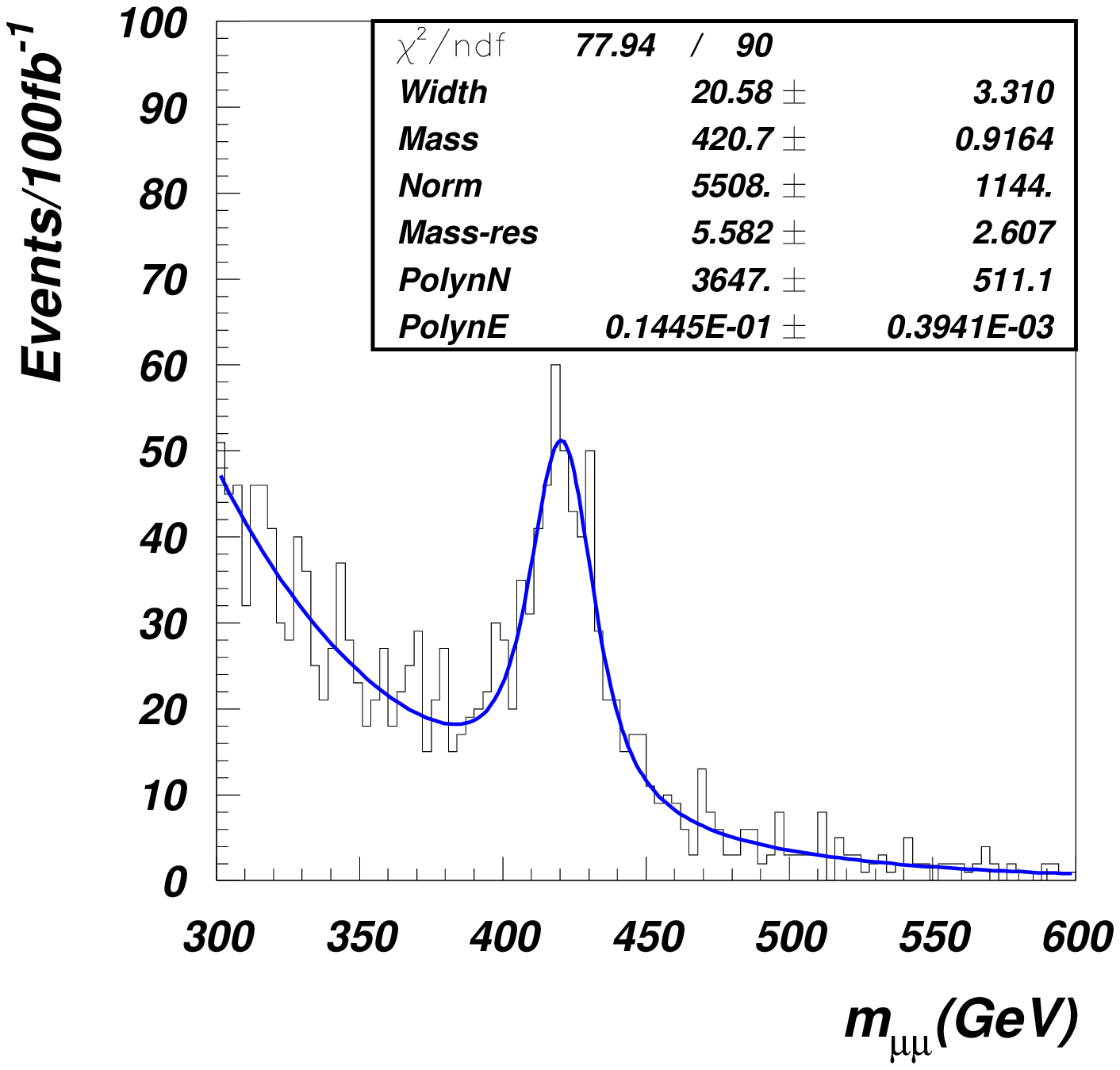,width=0.5\textwidth,height=0.25\textheight}%
\epsfig{file=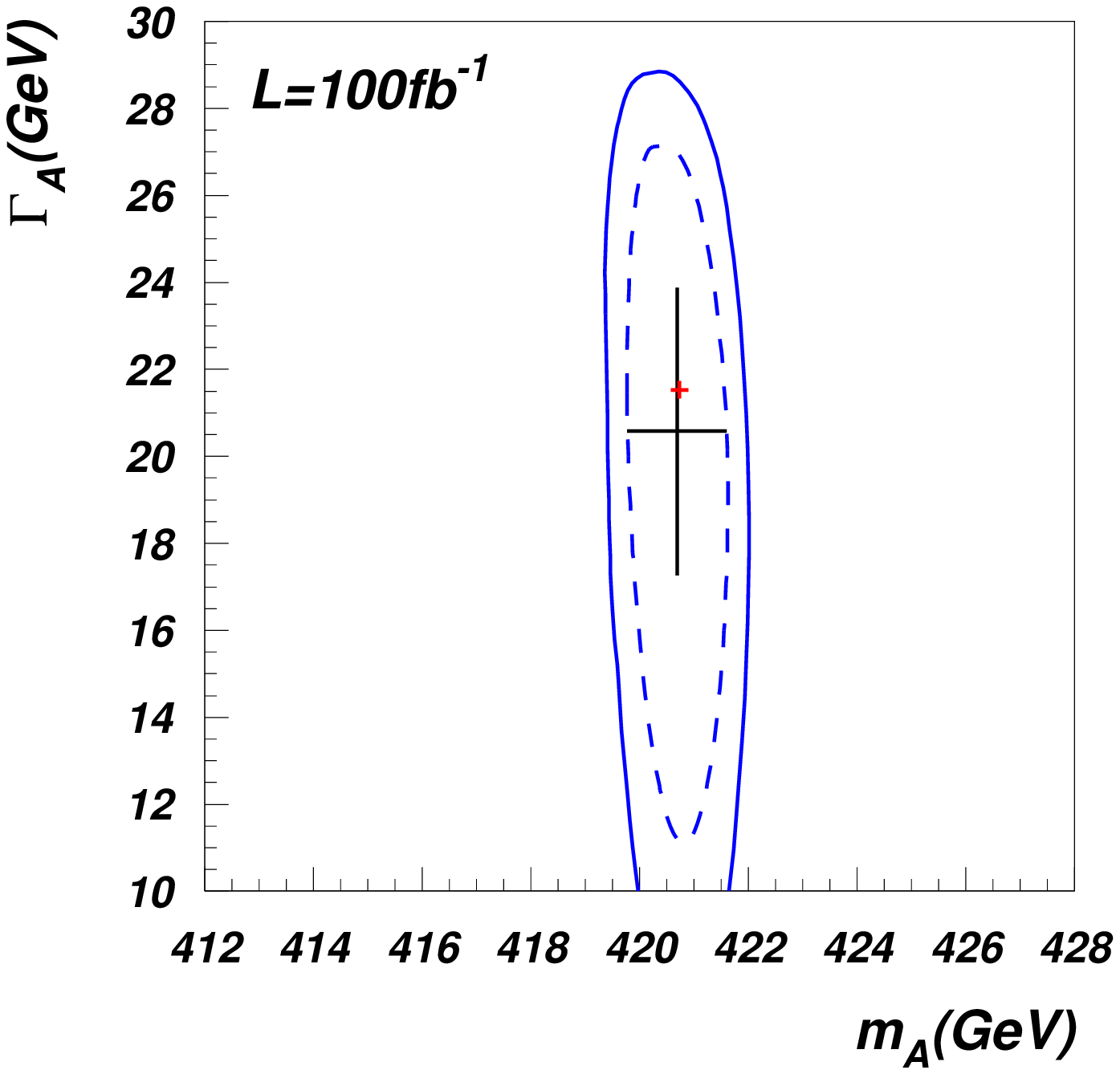,width=0.5\textwidth,height=0.25\textheight}\\
\vskip -1.0cm
\epsfig{file=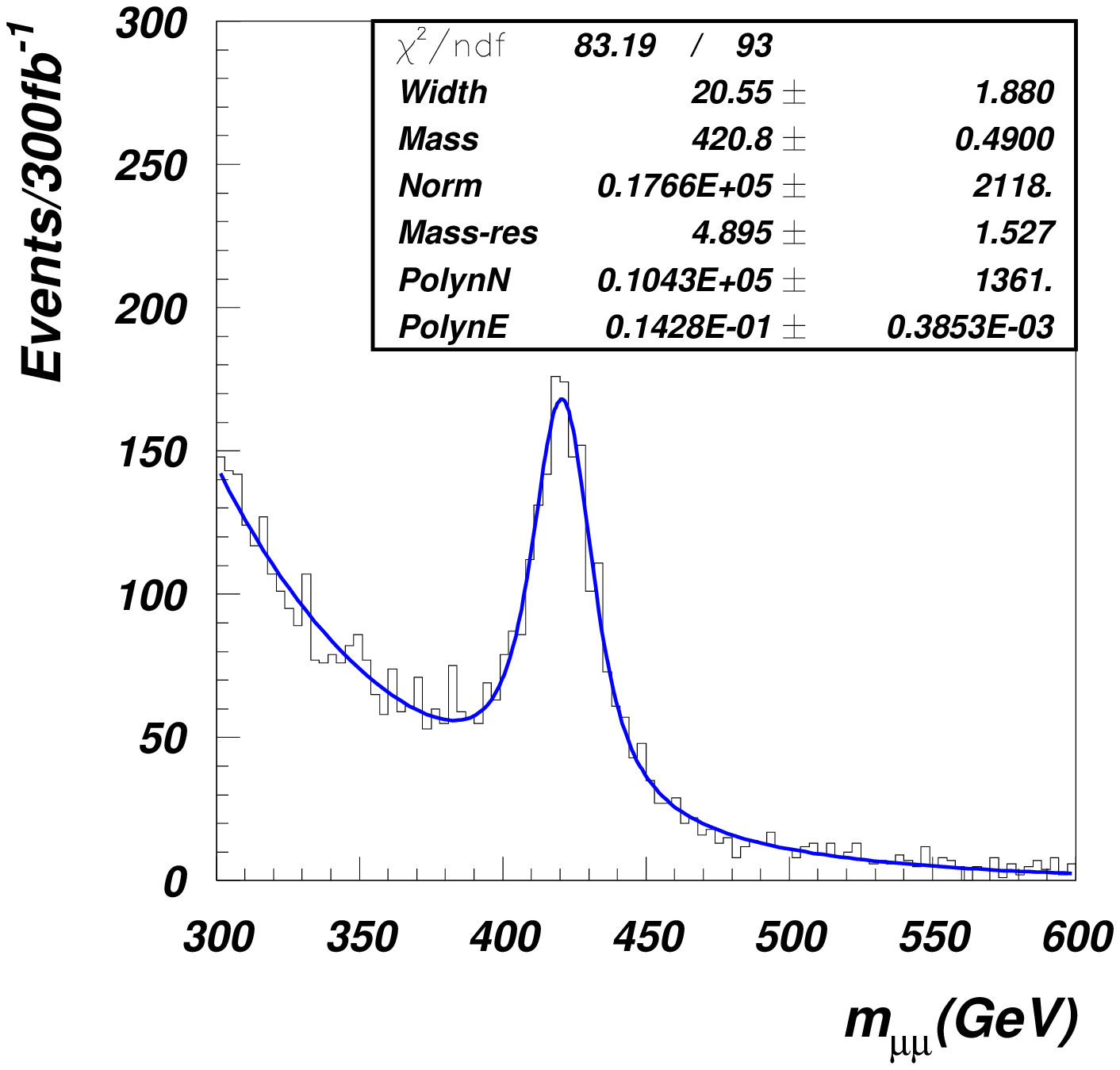,width=0.5\textwidth,height=0.25\textheight}%
\epsfig{file=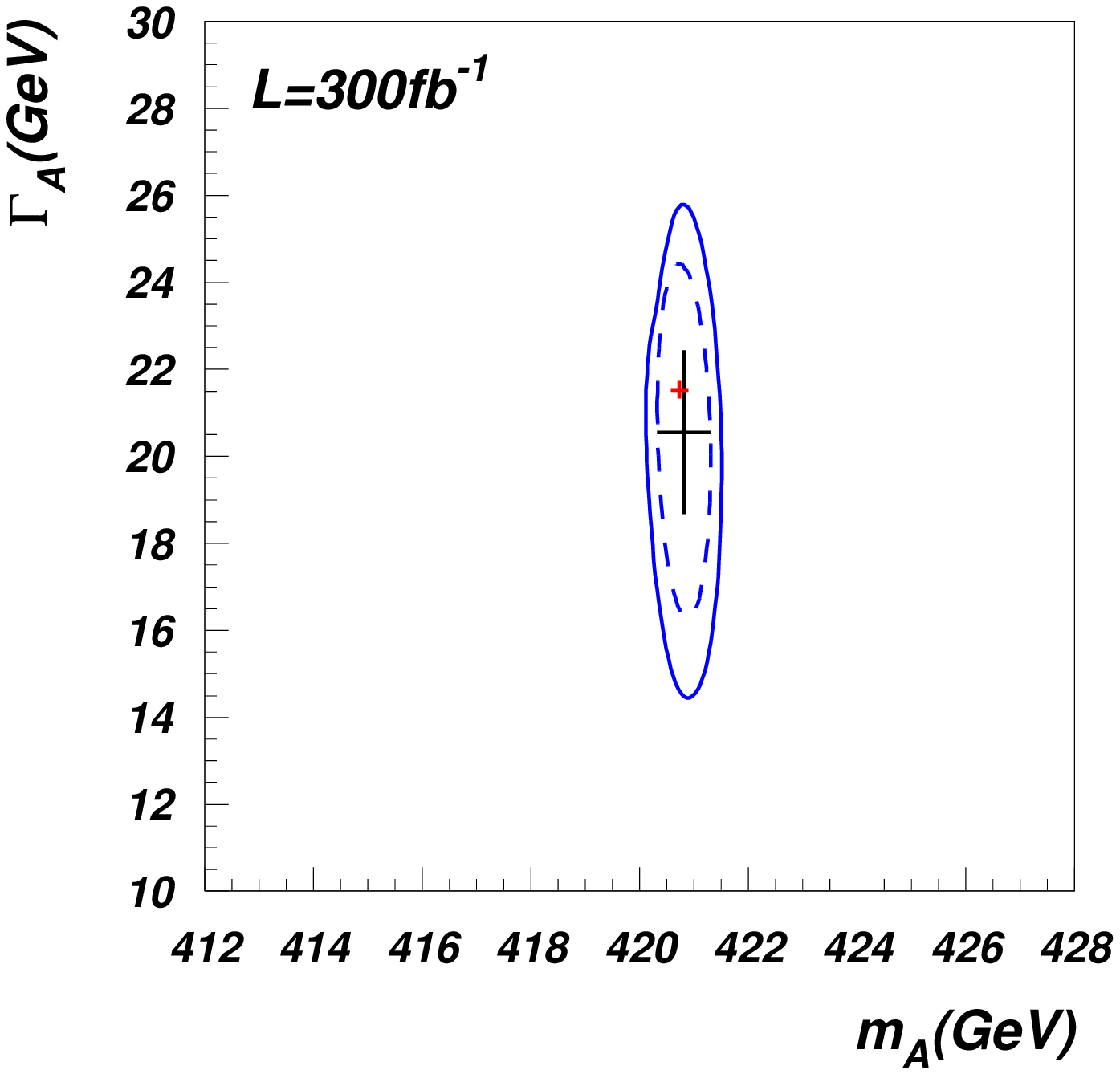,width=0.5\textwidth,height=0.25\textheight}\\
\vskip -1.0cm
\epsfig{file=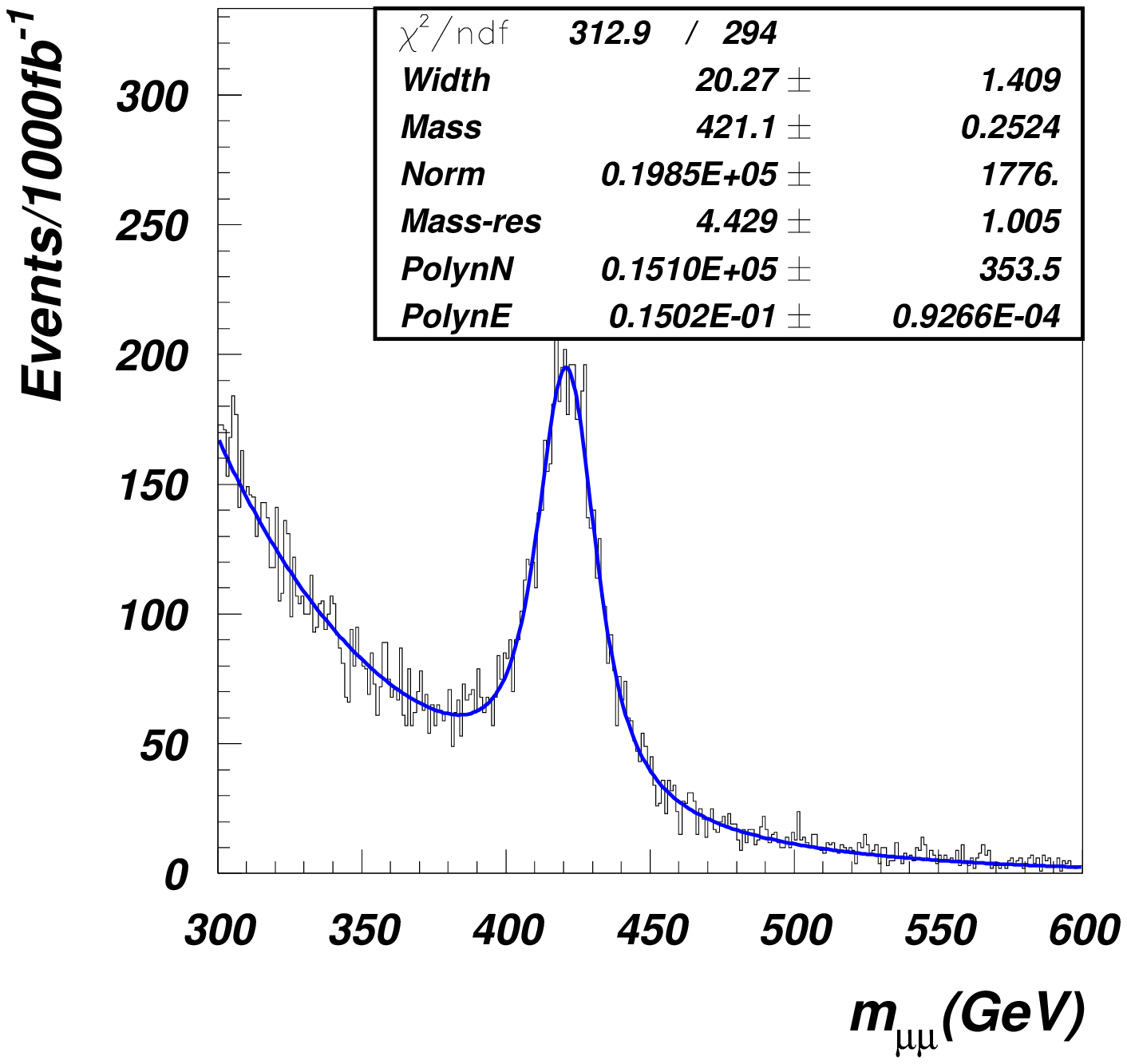,width=0.5\textwidth,height=0.25\textheight}%
\epsfig{file=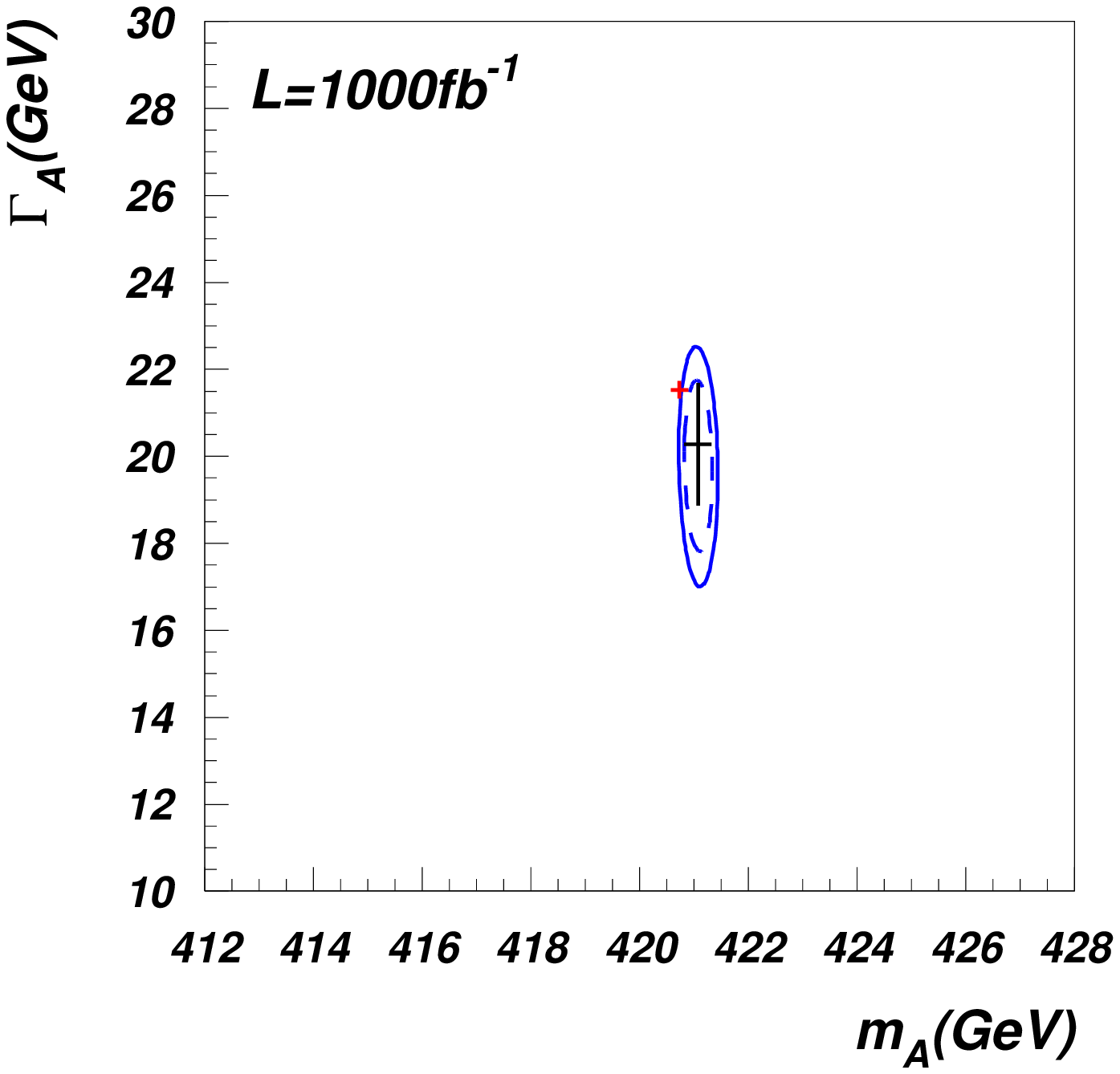,width=0.5\textwidth,height=0.25\textheight}\\
\vskip -1.0cm
\caption{\it 
Left :
best fit of Monte Carlo data for LCC4 for $pp\to bA,H\to b\mu^+\mu^- +X$
production including muon smearing.
Right: corresponding contours of fit to $m_A$ and $\Gamma _A$ values for 
Monte Carlo data for LCC4 from $pp\to bA,H\to b\mu^+\mu^- +X$
production including muon smearing.
}\label{fig:fit_smear} 
\end{figure}

The results from the $\chi^2$ fits of signal-plus-background are presented in 
Fig.~\ref{fig:fit_smear} for different integrated luminosities of $L=$30, 100, 300 and 1000 fb$^{-1}$.
The left side  of the Fig.~\ref{fig:fit_smear} shows the
fit to Monte Carlo data for LCC4 with $pp\to bA,H\to b\mu^+\mu^- +X$
production including muon smearing.
It also shows the  values of the fitted parameters
$(\Gamma, M, N, \sigma, N_{p1},N_{p2})$ together 
with their standard deviations according to the fit.
The fit has been performed using the MINUIT program from
CERN library which properly takes into account the 
correlation matrix of the fit parameters, which
is crucial for the evaluation of the 
corresponding contours  in the $\Gamma_A$ {\it vs.} $m_A$ plane
at $1\sigma$ and $2\sigma$ confidence levels; these are
shown on the right side of the Fig.~\ref{fig:fit_smear}.
The black crosses show the width measurement assuming the Higgs masses are known
to perfect accuracy.
\begin{figure}[htbp]
\epsfig{file=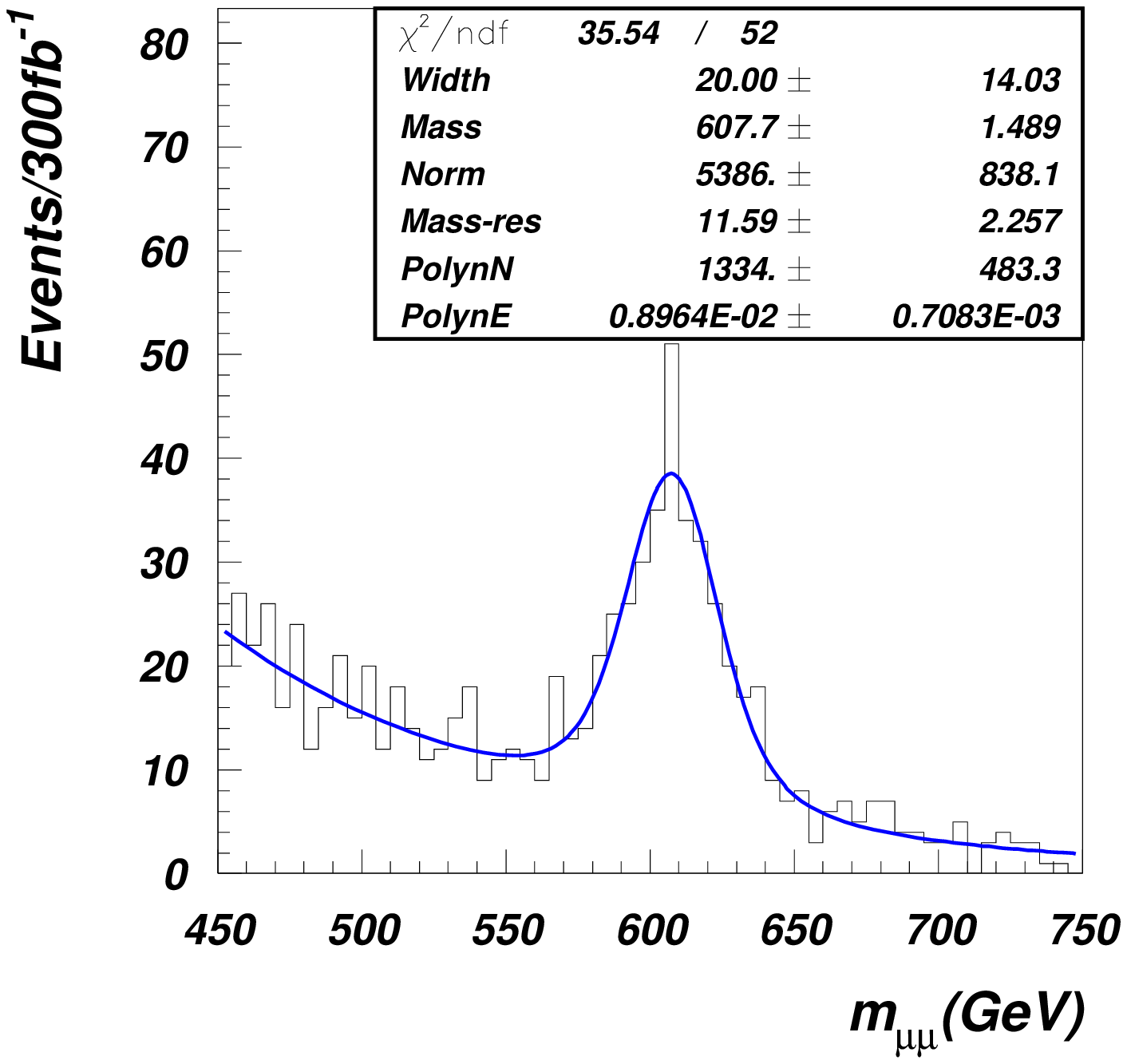,width=0.5\textwidth,height=0.25\textheight}%
\epsfig{file=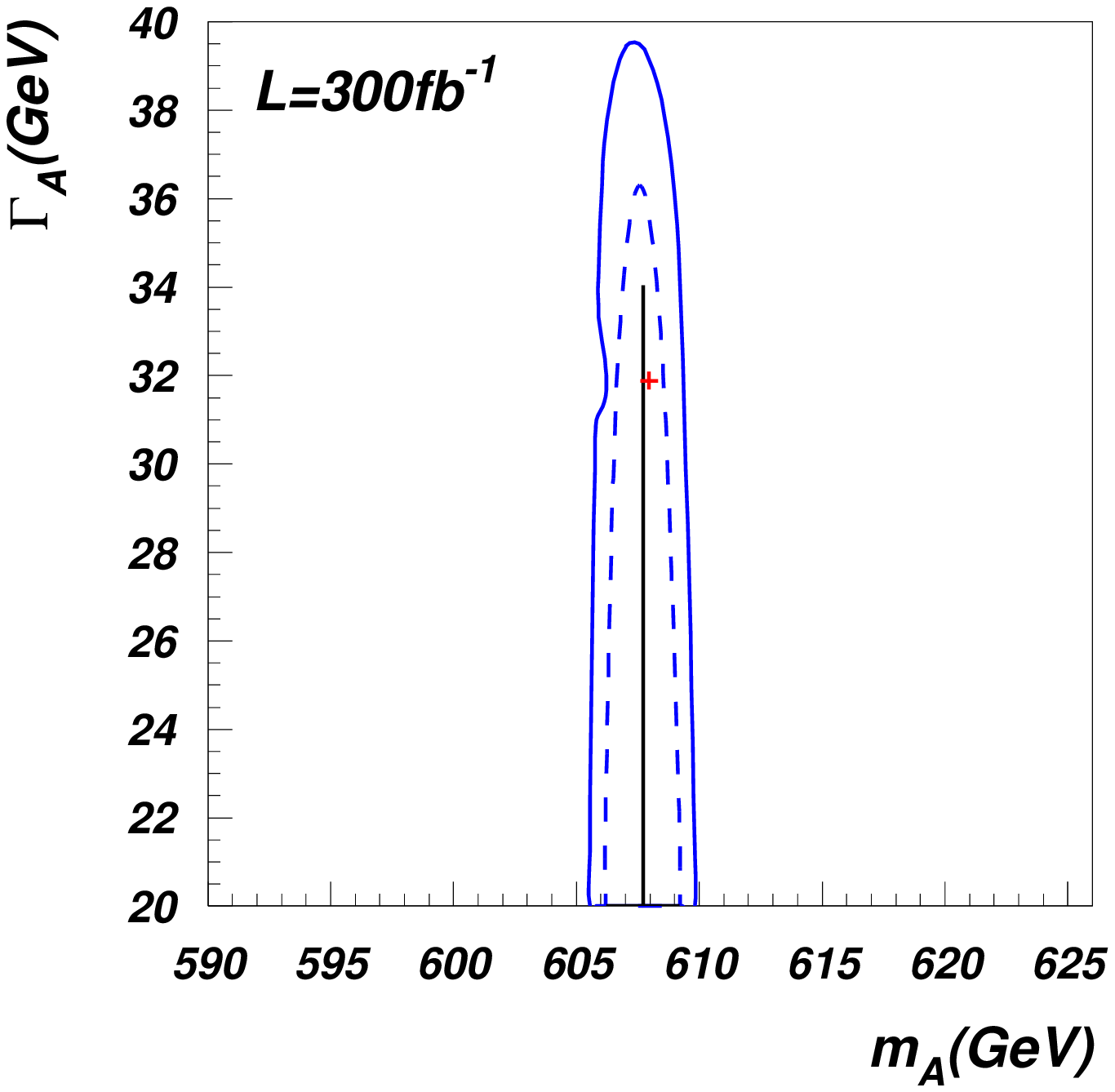,width=0.5\textwidth,height=0.25\textheight}\\
\vskip -1.0cm
\epsfig{file=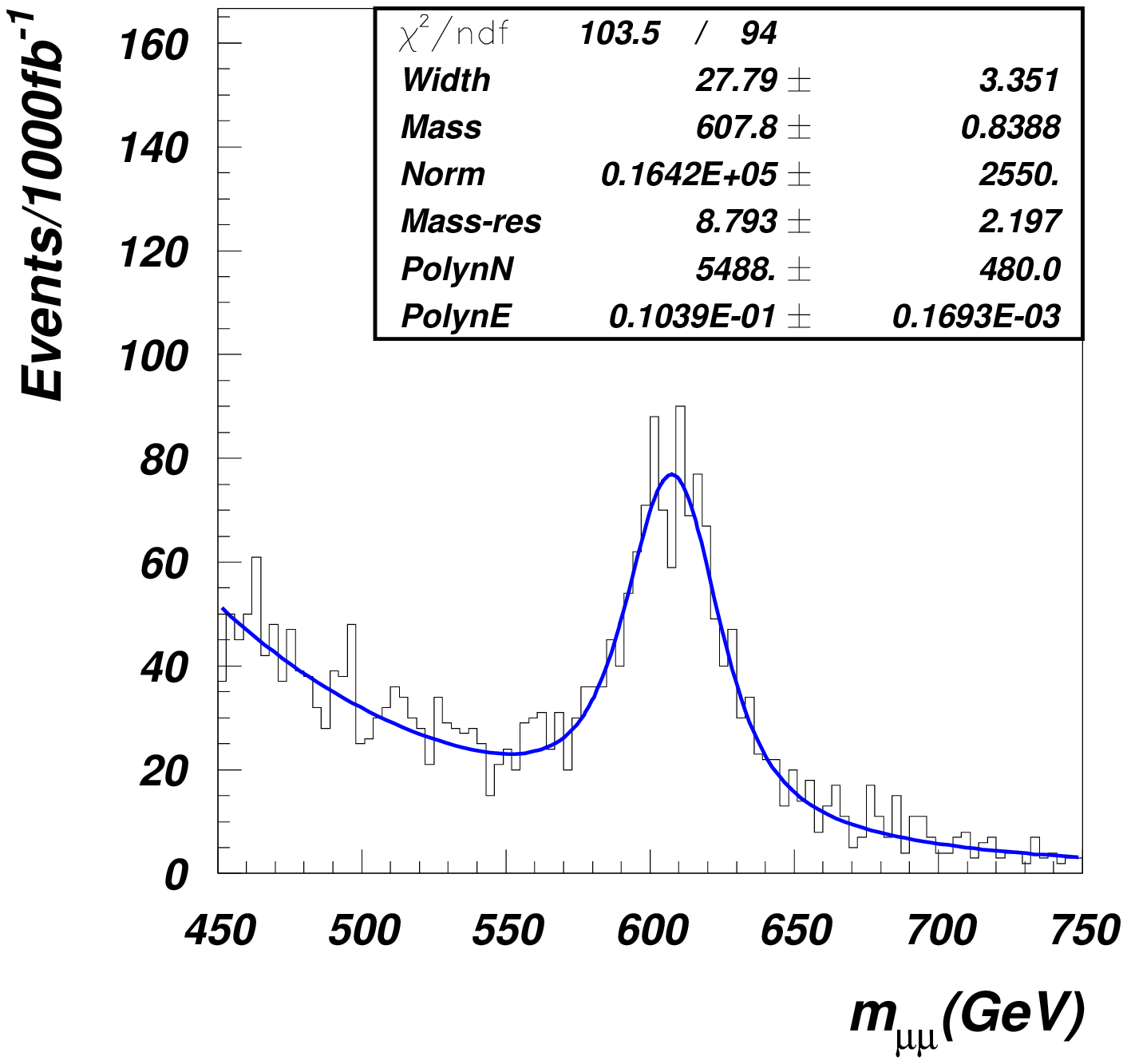,width=0.5\textwidth,height=0.25\textheight}%
\epsfig{file=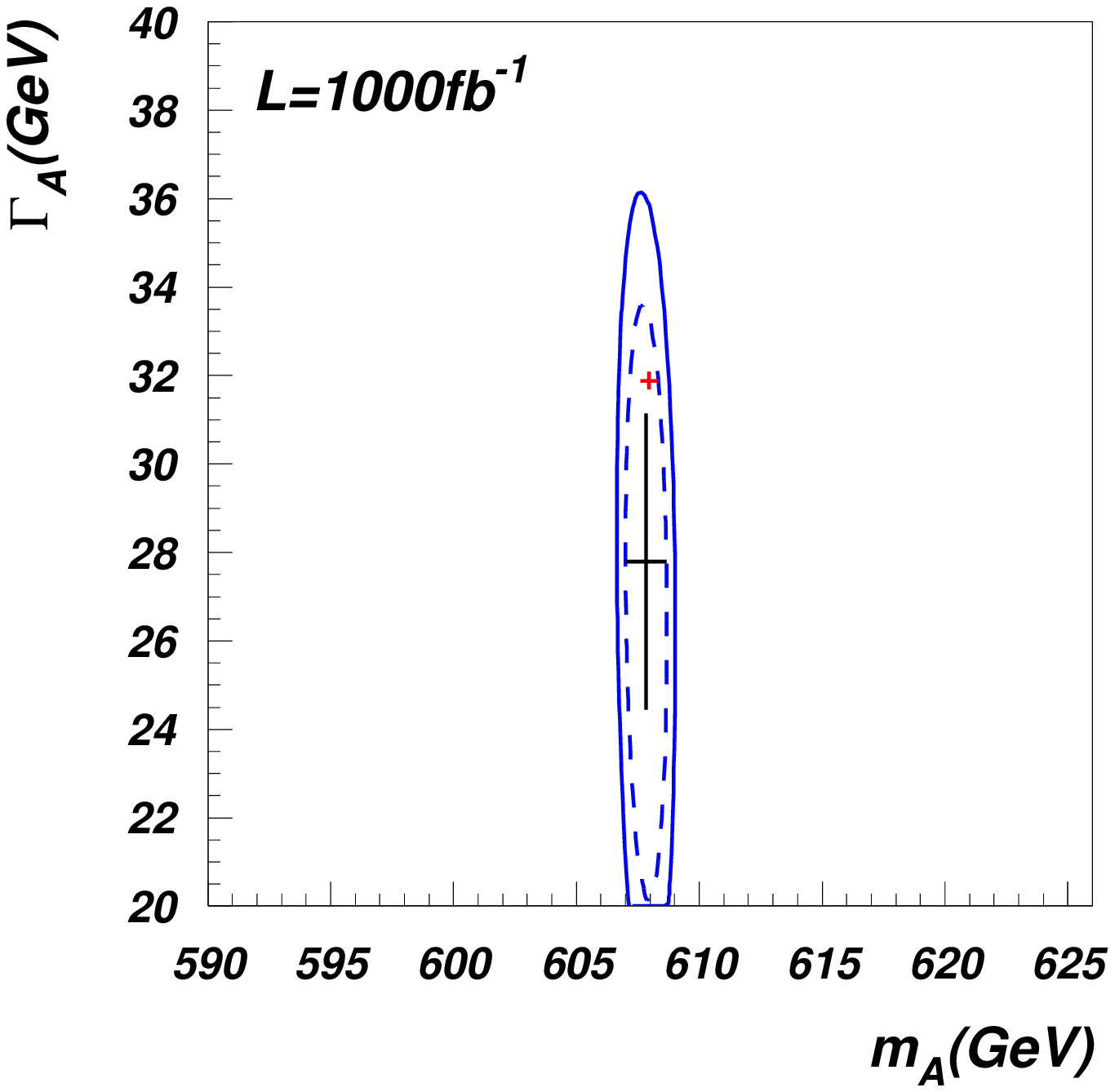,width=0.5\textwidth,height=0.25\textheight}\\
\vskip -1.0cm
\caption{\it 
Left :
best fit of Monte Carlo data for BM600 for $pp\to bA,H\to b\mu^+\mu^- +X$
production including muon smearing.
Right: corresponding contours of fit to $m_A$ and $\Gamma _A$ values for 
Monte Carlo data for BM600 from $pp\to bA,H\to b\mu^+\mu^- +X$
production including muon smearing.
}\label{fig:fit_600} 
\end{figure}
From Fig. \ref{fig:fit_smear}, one can see that for $L=$30 fb$^{-1}$
of data, the statistics only provide a rough fit to the $A,\ H$ width. 
On the other hand, moving to $L=$100 fb$^{-1}$, 
our fit provides promising results for $\Gamma_A$. We see that with $L=$100 fb$^{-1}$, 
$m_A$ can be measured to 1 GeV accuracy, or $0.25\%$.
Meanwhile, the $A,\ H$ width is measured at $\Gamma_{A,H}\simeq 20\pm 8$ GeV, 
or 40\% level.
At higher integrated luminosity values of $L=$300 fb$^{-1}$, 
the accuracy on  $m_{A,H}$ is improved to sub-GeV levels and 
$\Gamma_{A,H}$ is found to be $\sim 20\pm 4$ GeV, a 20\% measurement.
At  $L=$1000 fb$^{-1}$, which might be reached in $\sim 10$  years of 
LHC running, the measurement of $\Gamma_{A,H}$ can be improved to about 
$20\pm 1.75$ GeV, or $\sim 8\%$ accuracy.
The $\Gamma_{A,H}$ accuracy is expected to approach $\sim 7\%$ level for 
infinite integrated luminosity, and is mainly limited by the detector muon 
energy resolution of $4\%$,
which is actually quite close to $\Gamma_A/m_A$.

The results from the $\chi^2$ fits to $m_A$ and $\Gamma_A$ of signal-plus-background 
for benchmark point BM600 are shown in 
Fig.~\ref{fig:fit_600} for integrated luminosities of $L=$300 and 1000 fb$^{-1}$.
In this case, the heavy Higgs masses are $m_{A,H}=608,\ 612$ GeV, while the widths
from CalcHEP are $\Gamma_{A,H}=31.9\ (32.1)$ GeV.
We find for $10^3$ fb$^{-1}$ of integrated luminosity, that $\Gamma_A$ is 
extracted to be $\Gamma_A=28\pm 5.5$ GeV, a $17\%$ measurement. 

\section{Conclusions}
\label{sec:conclude}

In $R$-parity-conserving supersymmetric models where the lightest neutralino 
is expected to be a thermal relic of the Big Bang, and to comprise the dark 
matter in the universe, special qualities are needed to enhance the dark 
matter annihilation rates. One compelling case is neutralino annihilation 
through the pseudoscalar Higgs resonance. 
Can the LHC tell us if such a mechanism is operative in the early universe?
The crucial test here is whether the condition $2m_{\tz_1}\sim m_A$ is 
fulfilled.

A variety of techniques have been proposed for extracting the SUSY particle 
masses --- including $m_{\tz_1}$ --- in sparticle cascade decay events 
at the LHC~\cite{masses}.
Extracting the heavy Higgs masses is also possible provided that $m_A$ 
is small enough and that $\tan\beta$ is large enough. Mass measurements 
of heavy Higgs decays into $b\bar{b}$ and $\tau^+\tau^-$ are fraught 
with uncertainties from multi-particle production and energy loss from 
neutrinos. We focused instead on the suppressed decay 
$A,\ H\to \mu^+\mu^-$, 
since it allows for both highly accurate heavy Higgs mass and width 
reconstructions. Production of $A$ and $H$ in association with 
a single $b$-jet offers a large background rejection at small cost to signal, 
especially in the large $\tan\beta$ regime, where Higgs production
in association with $b$s is expected to be enhanced by large Yukawa couplings.
This is also the regime in models such as mSUGRA where neutralino 
annihilation through the heavy Higgs resonance is expected to occur.

In this paper, we have computed regions of $m_A\ vs.\ \tan\beta$ parameter 
space where $pp\to b\phi +X, \phi = H,A$ production followed by 
$H,\ A\to\mu^+\mu^-$ should be visible for various integrated luminosities. 
We have also performed detailed Monte Carlo simulations of signal 
and background for the LCC4 and BM600 benchmark points. 
Fits of the dimuon mass spectra 
allow for sub-percent determinations of the (nearly overlapping) $H$ 
and $A$ masses. The $A,\ H$ overlapping widths were determined to 
$\sim 8\%$ ($\sim 17\%$) accuracy in the case of LCC4 (BM600) 
with $10^3$ fb$^{-1}$ of integrated luminosity. 
We conclude that indeed the study of $pp\to b\phi +X, \phi = H,A$ production followed by 
$H,\ A\to\mu^+\mu^-$  offers a unique opportunity to directly
measure $A(H)$ Higgs width. This process also allows to measure the $A\ (H)$
mass with unprecedented precision. Both these measurements would provide 
crucial information to connect the cosmological $A-funnel$ scenario 
of dark matter annihilation with LHC data.
Combining these measurements with SUSY particle mass measurements
such as the mass edge in $m_{\ell^+\ell^- }$ from $\tz_2\to\tz_1\ell^+\ell^-$ decay 
would go a long way towards determining the parameter $\tan\beta$, 
and also whether or not neutralino annihilation through the $A$ resonance
(with $2m_{\tz_1}\sim m_A$)
is the operative mechanism in the early universe to yield the measured 
abundance of neutralino dark matter.

\section*{Acknowledgements} 

This research was supported in part by the U.S. Department of Energy
under Grant No.~DE-FG02-04ER41305
and  Royal Society Grant No. 508786101.
PS thanks NExT Institute as a part of SEPnet for  support.



\begin{thebibliography}{99}
\small
%
\bibitem{wimp} For reviews, see {\it e.g.}
C.~Jungman, M.~Kamionkowski and K.~Griest,\prep{267}{1996}{195};
A.~Lahanas, N.~Mavromatos and D.~Nanopoulos, \ijmpd{12}{2003}{1529};
M.~Drees, \hepph{0410113};
K.~Olive, ``Tasi Lectures on Astroparticle Physics'', \astroph{0503065};
G. Bertone, D. Hooper and J. Silk, \prep{405}{2005}{279}.
%
\bibitem{bbox} H. Baer and A. Box, \epjc{68}{2010}{523};
H. Baer, A. Box and H. Summy, \jhep{1010}{2010}{023}.
%
\bibitem{wmap7} E. Komatsu {\it et al.} (WMAP collaboration), 
arXiv:1001.4538 (2010).
%
\bibitem{href} A.~G.~Riess {\it et al.},
  Astrophys.\ J.\  {\bf 699} (2009) 539 and arXiv:1103.2976 (2011).
%
\bibitem{stau} J.~Ellis, T.~Falk and K.~Olive, \plb{444}{1998}{367}; 
J.~Ellis, T.~Falk, K.~Olive and M.~Srednicki, \app{13}{2000}{181};
M.E.~G\'{o}mez, G.~Lazarides and C.~Pallis, \prd{61}{2000}{123512}
and \plb{487}{2000}{313};
A.~Lahanas, D.~V.~Nanopoulos and V.~Spanos, \prd{62}{2000}{023515};
R.~Arnowitt, B.~Dutta and Y.~Santoso, \npb{606}{2001}{59}; 
see also Ref.~\cite{isared}.
%
\bibitem{stop} C.~B\"ohm, A.~Djouadi and M.~Drees,
  \prd{30}{2000}{035012}; 
J.~R.~Ellis, K.~A.~Olive and Y.~Santoso, \app{18}{2003}{395};
J.~Edsj\"o, {\it et al.}, JCAP {\bf 0304} (2003) 001
%
\bibitem{ino} K. Griest and D. Seckel, \prd{43}{1991}{3191};
J. Edsjo and P. Gondolo, \prd{56}{1997}{1879};
J.~Edsjo, M.~Schelke, P.~Ullio and P.~Gondolo,
  JCAP {\bf 0304} (2003) 001
  [arXiv:hep-ph/0301106].

%
\bibitem{wtn} N. Arkani-Hamed, A. Delgado and G. Giudice, 
\npb{741}{2006}{108};
H.~Baer, A.~Mustafayev, E.~Park and X.~Tata,
JCAP{\bf 0701}, 017 (2007) and \jhep{0805}{2008}{058}.
%
\bibitem{hb_fp} K.~L.~Chan, U.~Chattopadhyay and P.~Nath, \prd{58}{1998}{096004};
J.~Feng, K.~Matchev and T.~Moroi, \prl{84}{2000}{2322} and 
\prd{61}{2000}{075005}; see also 
H.~Baer, C.~H.~Chen, F.~Paige and X.~Tata, \prd{52}{1995}{2746} and 
\prd{53}{1996}{6241}; 
H.~Baer, C.~H.~Chen, M.~Drees, F.~Paige and X.~Tata, \prd{59}{1999}{055014}; 
for a model-independent approach, see
H.~Baer, T.~Krupovnickas, S.~Profumo and P.~Ullio, \jhep{0510}{2005}{020}.
%
\bibitem{Afunnel} M.~Drees and M.~Nojiri, \prd{47}{1993}{376}; 
H.~Baer and M.~Brhlik, \prd{57}{1998}{567};
H.~Baer, M.~Brhlik, M.~Diaz, J.~Ferrandis, P.~Mercadante,
P.~Quintana and X.~Tata, \prd{63}{2001}{015007};
J.~Ellis, T.~Falk, G.~Ganis, K.~Olive and M.~Srednicki, \plb{510}{2001}{236}; 
V.~D.~Barger and C.~Kao,
Phys.\ Lett.\  B {\bf 518} (2001) 117; 
L.~Roszkowski, R.~Ruiz de Austri and T.~Nihei, \jhep{0108}{2001}{024}; 
A.~Djouadi, M.~Drees and J.~L.~Kneur, \jhep{0108}{2001}{055}; 
A.~Lahanas and V.~Spanos, \epjc{23}{2002}{185}.
%
\bibitem{nuhm} H.~Baer, A.~Mustafayev, S.~Profumo, A.~Belyaev and X.~Tata,
\prd{71}{2005}{095008}; 
H.~Baer, A.~Mustafayev, S.~Profumo, A.~Belyaev and X.~Tata, 
\jhep{0507}{2005}{065}.
%
\bibitem{ltanb} H. Baer, C. H. Chen, M. Drees, F. Paige and X. Tata,
\prl{79}{1997}{986}.
%
\bibitem{willen} J. Campbell, R. K. Ellis, F. Maltoni and S. Willenbrock, 
\prd{67}{2003}{095002}.
%
\bibitem{lhcHiggsreach} 
C.~Kao, D.~A.~Dicus, R.~Malhotra and Y.~Wang,
  Phys.\ Rev.\  D {\bf 77} (2008) 095002;
  C.~Kao, S.~Sachithanandam, J.~Sayre and Y.~Wang,
  Phys.\ Lett.\  B {\bf 682} (2009) 291.
%
\bibitem{stepanov}
  C.~Kao, N.~Stepanov, Phys.\ Rev.\  {\bf D52}, 5025-5030 (1995).
\bibitem{ddkm} S. Dawson, D. Dicus, C. Kao and R. Malhotra, 
\prl{92}{2004}{241801}; 
C.~Kao and Y.~Wang,
  Phys.\ Lett.\  B {\bf 635} (2006) 30.
%
\bibitem{isared} IsaReD, by H. Baer, C. Balazs and A.Belyaev, \jhep{0203}{2002}{042}.
%
\bibitem{isajet} ISAJET v7.79, by H. Baer, F. Paige, S. Protopopescu and
X. Tata, \hepph{0312045}; for details on the Isajet spectrum calculation, see
H. Baer, J. Ferrandis, S. Kraml and W. Porod, \prd{73}{2006}{015010}.
%
\bibitem{bbbkt} H. Baer, C. Balazs, A. Belyaev, T. Krupovnickas and X. Tata,
\jhep{0306}{2003}{054}
%
\bibitem{cascade} H. Baer, J. Ellis, G. Gelmini, D. V. Nanopoulos
and X. Tata, \plb{161}{1985}{175}; 
G. Gamberini, \zpc{30}{1986}{605}; H. Baer, V. Barger, 
D. Karatas and X. Tata, \prd{36}{1987}{96}; H. Baer, X. Tata and
J. Woodside, \prd{45}{1992}{142}.
%
\bibitem{ofarril} H. Baer and J. O'Farrill, JCAP{\bf 0404}, 005 (2004);
H.~Baer, A.~Belyaev, T.~Krupovnickas and J.~O'Farrill, 
JCAP {\bf 0408} (2004) 005;
H. Baer, E. K. Park and X. Tata, \njp{11}{2009}{105024}.
%
\bibitem{xe100} E. Aprile {\it et al} (Xenon100 Collaboration), 
\prl{105}{2010}{131302}.
%
\bibitem{mlledge} H. Baer, K. Hagiwara and X. Tata, \prd{35}{1987}{1598};
H. Baer, D. Dzialo-Karatas and X. Tata, \prd{42}{1990}{2259};
H. Baer, C. Kao and X. Tata, \prd{48}{1993}{5175};
H. Baer, C. H. Chen, F. Paige and X. Tata, \prd{50}{1994}{4508};
I.~Hinchliffe {\it et al.}, \prd{55}{1997}{5520}
and \prd{60}{1999}{095002}; H.~Bachacou, I.~Hinchliffe and F.~Paige, 
\prd{62}{2000}{015009}.


\bibitem{Kretzer:2003it}
  S.~Kretzer, H.~L.~Lai, F.~I.~Olness and W.~K.~Tung,
  Phys.\ Rev.\  D {\bf 69}, 114005 (2004)
  [arXiv:hep-ph/0307022].




%
\bibitem{pbmz} D. Pierce, J. Bagger, K. Matchev and R. Zhang, 
\npb{491}{1997}{3}.
%
\bibitem{HtoSUSY} H. Baer, M. Bisset, D. Dicus, C. Kao  and X. Tata,
\prd{47}{1993}{1062};
H. Baer, M. Bisset, C. Kao  and X. Tata,
\prd{50}{1994}{316}.
%
\bibitem{peskin} E. Baltz, M. Battaglia, M. Peskin and T. Wizansky,
\prd{74}{2006}{103521}.
%
\bibitem{Bayatian:2006zz}
  G.~L.~Bayatian {\it et al.}  [CMS Collaboration],

\bibitem{Ball:2007zza}
  G.~L.~Bayatian {\it et al.}  [CMS Collaboration],
  J.\ Phys.\ G {\bf 34}, 995 (2007).


\bibitem{calchep}  CalcHEP, by A. Pukhov, \hepph{0412191}.
%
\bibitem{masses} For a review, see {\it e.g.} A.~J.~Barr and C.~G.~Lester,
  J.\ Phys.\ G {\bf 37} (2010) 123001.
%
%
%
%
\end{thebibliography}
\end{document}